\documentclass[aps,prd,twocolumn,preprintnumbers,superscriptaddress,nofootinbib]{revtex4-2}

\usepackage{amsmath,amssymb,graphicx,orcidlink,xcolor}
\usepackage[normalem]{ulem}

\usepackage{subcaption} 
\usepackage[dvipsnames]{xcolor}
\usepackage[font=small, labelfont=bf, justification=raggedright, singlelinecheck=false]{caption}

\newcommand{\dd}[0]{\mathrm{d}}
\newcommand{\iso}[2]{{}^{#1}\mathrm{#2}}

\definecolor{apsblue}{RGB}{0,0,128}
\hypersetup{
  colorlinks = true,
  allcolors  = apsblue
}

\begin{document}

\title{Probing the axion-nucleon coupling with supergiant stars}

\author{Francisco R. Cand\'on~\orcidlink{0009-0002-3199-9278}}
\affiliation{Fakult\"at f\"ur Physik, TU Dortmund, Otto-Hahn-Straße 4a, Dortmund D-44221, Germany}
\affiliation{CAPA \& Departamento de F\'isica Te\'orica, Universidad de Zaragoza, C. Pedro Cerbuna 12, 50009 Zaragoza, Spain}

\author{Pablo Casaseca~\orcidlink{0009-0009-5399-1569}}
\affiliation{CAPA \& Departamento de F\'isica Te\'orica, Universidad de Zaragoza, C. Pedro Cerbuna 12, 50009 Zaragoza, Spain}

\author{Maurizio Giannotti~\orcidlink{0000-0001-9823-6262}}
\affiliation{CAPA \& Departamento de F\'isica Te\'orica, Universidad de Zaragoza, C. Pedro Cerbuna 12, 50009 Zaragoza, Spain}

\author{Mathieu Kaltschmidt~\orcidlink{0000-0002-6470-5371}}

\affiliation{CAPA \& Departamento de F\'isica Te\'orica, Universidad de Zaragoza, C. Pedro Cerbuna 12, 50009 Zaragoza, Spain}

\author{Jaime Ruz~\orcidlink{0000-0002-3253-7027}}
\affiliation{Fakult\"at f\"ur Physik, TU Dortmund, Otto-Hahn-Straße 4a, Dortmund D-44221, Germany}
\affiliation{CAPA \& Departamento de F\'isica Te\'orica, Universidad de Zaragoza, C. Pedro Cerbuna 12, 50009 Zaragoza, Spain}

\author{Julia K. Vogel~\orcidlink{0000-0002-5850-5517}}
\affiliation{Fakult\"at f\"ur Physik, TU Dortmund, Otto-Hahn-Straße 4a, Dortmund D-44221, Germany}
\affiliation{CAPA \& Departamento de F\'isica Te\'orica, Universidad de Zaragoza, C. Pedro Cerbuna 12, 50009 Zaragoza, Spain}

\begin{abstract}
A finite axion-nucleon coupling enables the production of axions in stellar environments via the thermal
excitation and subsequent deexcitation of the $\iso{57}{Fe}$ isotope. 
Given its low-lying excited state at 14.4 keV, $\iso{57}{Fe}$ can be efficiently excited in the hot cores of supergiant stars, possibly leading to axion emission. 
The conversion of these axions into photons in the Galactic magnetic field results in a characteristic 14.4 keV line, potentially detectable by hard x-ray telescopes such as NASA’s Nuclear Spectroscopic Telescope Array (\textsc{NuSTAR}). 
In this work, we present the first constraints on axion-nucleon couplings derived from \textsc{NuSTAR} observations of Betelgeuse and discuss the potential insights that could be gained from detecting this line in other nearby supergiants. Our results establish significantly more stringent bounds than those obtained from solar observations, setting a limit of $|g_{a\gamma} g_{aN}^{\mathrm{eff}}| < (1.2 - 2.7) \times 10^{-20}$ GeV$^{-1}$ for $m_a \lesssim 10^{-10}$ eV.\\

\noindent DOI: \href{https://journals.aps.org/prd/abstract/10.1103/xv5c-j2rv}{10.1103/xv5c-j2rv}
\end{abstract}

\maketitle

\section{Introduction}
The QCD axion, often known just as axion, is a hypothetical pseudoscalar particle that emerges as an unavoidable consequence of the Peccei-Quinn solution of the strong $CP$ problem in QCD ~\cite{Weinberg:1977ma,Wilczek:1977pj,Peccei:1977hh,Peccei:1977ur}. 
In addition to the QCD
axion, many extensions of the Standard Model (SM),
particularly those inspired by string theory, predict the
existence of axionlike particles (ALPs), which share similar
properties but are not necessarily tied to the strong $CP$ problem \cite{Witten:1984dg,Conlon:2006tq,Arvanitaki:2009fg,Acharya:2010zx,Higaki:2011me,Cicoli:2012sz,Demirtas:2018akl,Mehta:2021pwf}. 
These particles are well-motivated candidates for new physics and can be probed through their weak interactions with SM particles.~\footnote{Throughout this paper, we use the terms “axion” and ALP
interchangeably. However, our analysis focuses on generic ALPs
rather than QCD axions. This choice is motivated by the ultralight
masses required for efficient photon-ALP conversion in the
Galactic magnetic field (see, e.g., Refs.~\cite{Xiao:2020pra,Dessert:2020lil}). Benchmark
QCD axion models predict extremely small couplings at these
masses ~\cite{DiLuzio:2020wdo}, placing them outside the reach of our study. As our
discussion focus on ALPs, we ignore any specific relation
between the mass and the different couplings. Nevertheless,
specific theoretical frameworks—such as the clockwork mechanism discussed in Ref.~\cite{Darme:2020gyx}—can yield viable combinations of
photon and nucleon couplings even within QCD axion models.}
A key feature of axions and ALPs is their coupling to photons, nucleons, and electrons, leading to a variety of potential experimental signatures. 
While most studies focus on the axion-photon interaction, due to its direct detectability in laboratory experiments and astrophysical observations \cite{Irastorza:2018dyq,DiVecchia:2019ejf,DiLuzio:2020wdo,Agrawal:2021dbo,Sikivie:2020zpn,Giannotti:2024xhx,Carenza:2024ehj}, the coupling to nucleons remains an important but challenging target. 
In QCD axion models, the nucleon coupling gets a contribution from the axion-gluon interaction, and its measurement could provide critical information about the axion’s underlying theory \cite{GrillidiCortona:2015jxo,DiLuzio:2020wdo}. 

However, direct experimental probes of this interaction are scarce, and most constraints arise from astrophysical and cosmological considerations \cite{Keller:2012yr,Sedrakian:2015krq,Hamaguchi:2018oqw,Beznogov:2018fda,Sedrakian:2018kdm,Leinson:2021ety,Turner:1987by,Burrows:1988ah,Raffelt:1987yt,Raffelt:1990yz,Carenza:2019pxu,Carenza:2020cis,Fischer:2021jfm,Lella:2022uwi,Lella:2023bfb,Lella:2024hfk}.

A promising venue to probe the axion-nucleon interaction is through nuclear transitions in astrophysical environments. 
As pseudoscalar particles, axions can be emitted in magnetic dipole (M1) nuclear transitions, a possibility already mentioned in the original axion proposal \cite{Weinberg:1977ma} and later developed, particularly in Ref.~\cite{Donnelly:1978ty}. One of the most studied cases is the 14.4 keV transition of $\iso{57}{Fe}$, which occurs when its first excited state decays to the ground state. 
Axions produced in this process can convert into x-ray photons in the presence of a Galactic or laboratory magnetic field via the Primakoff effect. 
This mechanism has been extensively studied in the Sun, where $\iso{57}{Fe}$ is relatively abundant, leading to constraints from dedicated axion searches such as CAST \cite{CAST:2009jdc}, CUORE \cite{Li:2015tyq}, and XENON1T \cite{XENON:2020rca}.
However, the Sun, with a core temperature of $\mathord{\approx}1$ keV, is not an optimal environment for axion searches via nuclear transitions. 
Supergiant stars, with core temperatures significantly higher than those of main-sequence (MS) stars like the Sun, offer a compelling alternative. 
In fact, the axion emission rate in nuclear deexcitations depends steeply on the temperature, making hotter stellar environments much more efficient sources. 
A particularly interesting case is the red supergiant Betelgeuse, located in the constellation of Orion at approximately 200 pc from Earth. 
This target is especially promising because observational data from the hard x-ray telescope aboard the Nuclear Spectroscopic Telescope Array (\textsc{NuSTAR}) already exists~\cite{Xiao:2020pra}, providing an opportunity to search for an axion-induced 14.4 keV line. 
Since this signature would appear as a monochromatic feature rather than a broad continuum, it offers a key observational advantage, as it is easier to distinguish from astrophysical backgrounds.
Furthermore, its detection could provide crucial insights into stellar properties. 
In particular, if the line could be resolved in future observations, it could offer a new probe of the core temperature as well as information on the rotational dynamics of the core, a quantity otherwise difficult to measure directly.

\begin{figure}[h]
    \centering
    \includegraphics[width=1.\columnwidth]{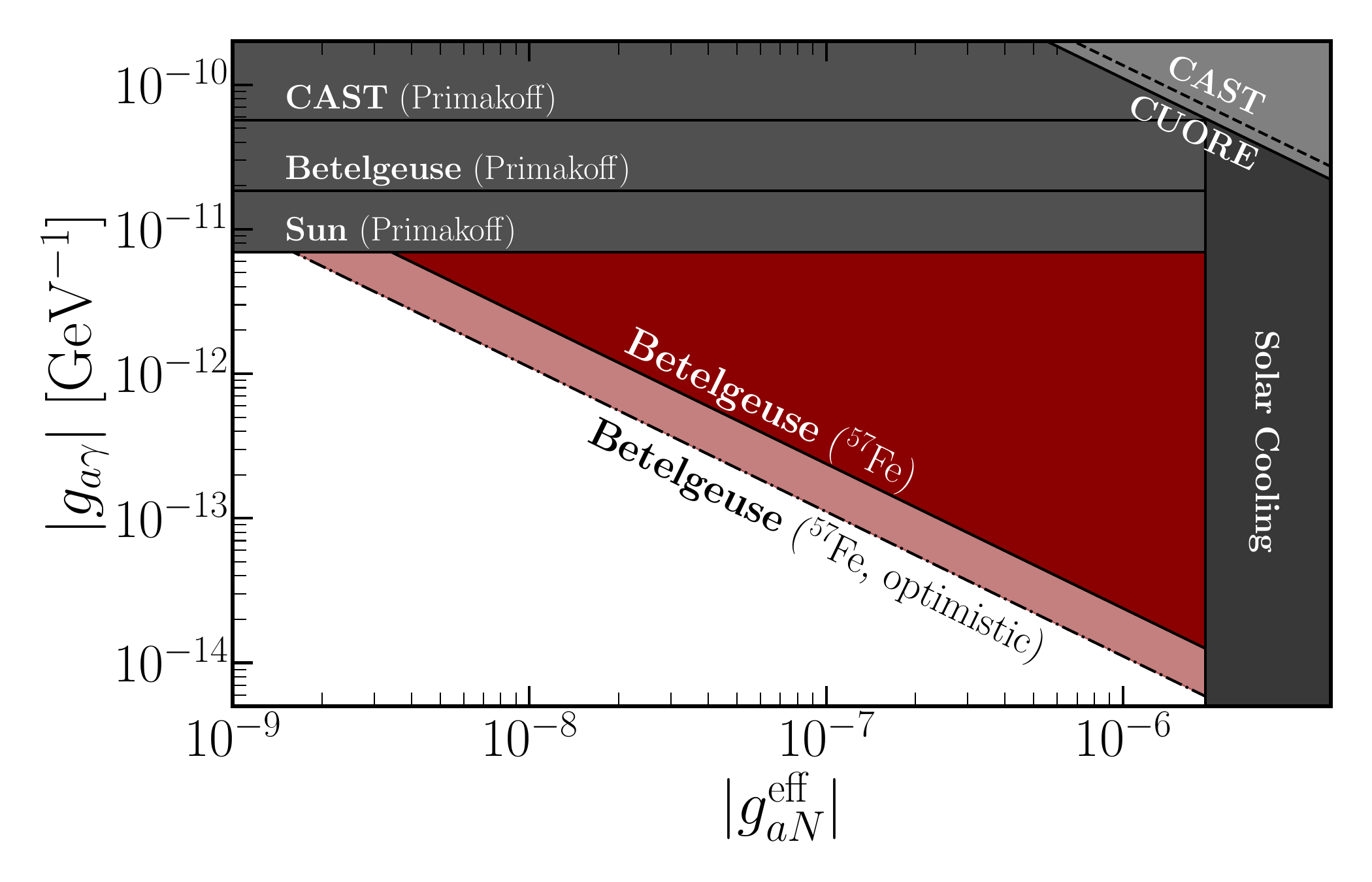}
    \caption{Exclusion limits from the Bayesian unbinned likelihood analysis at 95\% confidence level (CL), in the case of massless axions  compared to previous $^{57}$Fe studies. Our conservative results ($B_T = 1.4\ \mu G$) are displayed in dark red and a more optimistic bound, using $B_T = 3.0\ \mu G$, in a lighter red. Shown are the latest bounds on the axion-photon coupling $|g_{a\gamma}|$ from CAST \cite{CAST:2024eil} (dashed) and results from the Nuclear Spectroscopic Telescope Array (\textsc{NuSTAR}) observations of Betelgeuse \cite{Xiao:2020pra} and the Sun \cite{Ruz:2024gkl} in gray,  the solar cooling bound on the effective axion-nucleon coupling $|g_{aN}^{\mathrm{eff}}|$ \cite{DiLuzio:2021qct} in darker gray and results of searches for the $14.4$ keV line of $^{57}$Fe similar to this work from CAST \cite{CAST:2009jdc} (dashed) and CUORE \cite{Li:2015tyq} in a lighter gray.} 
    \label{fig:results_exclusion-massive}
\end{figure}

In this work, we present the first experimental axion search from $\iso{57}{Fe}$ nuclear decay in a star other than the Sun. While our analysis does not find any significant evidence for nonstandard physics, it allows us to set the strongest experimental bound on the axion-nucleon coupling for axion masses below $\mathord{\sim}10^{-10}$ eV. The results, shown in Fig.~\ref{fig:results_exclusion-massive}, demonstrate that our constraints are orders of magnitude stronger than those obtained from previous experimental searches.
Similar studies with other nearby supergiants are possible and thus future observations could further improve these bounds.
\section{Axion flux from Nuclear Transitions in Supergiants}
The axion production in stellar environments has been extensively studied and reviewed in the last decades~\cite{Raffelt:1996wa,Giannotti:2015kwo,Giannotti:2017hny,DiLuzio:2020jjp,DiLuzio:2021ysg,Caputo:2024oqc,Carenza:2024ehj}.
This production is mediated by the axion interaction with SM fields, described by the Lagrangian
\begin{equation}
    \mathcal{L}_{\mathrm{int}} = -\frac{1}{4}g_{a\gamma}aF_{\mu\nu}\tilde{F}^{\mu\nu} - \sum_i g_{af} a \bar{f}\gamma_5 f,
    \label{eqn:lagrangian}
\end{equation}
where $F$ is the electromagnetic field strength, $\tilde F$ its dual, and $f$ the fermion fields. 

Axions produced in nuclear fusion and decay processes typically carry energies on the order of $\mathord{\sim}$MeV. For instance, the $p+d \to {^{3}{\rm He}}+a$ reaction, which generates an axion flux at 5.5~MeV, has been experimentally investigated by Borexino~\cite{Borexino:2012guz}, CAST~\cite{CAST:2009klq}, and more recently analyzed using SNO data~\cite{Bhusal:2020bvx}. The potential for detecting such axions in next-generation experiments has also been explored in~\cite{Lucente:2022esm,Arias-Aragon:2024gdz}. However, the flux of axions from these reactions in the Sun is typically low, with an estimated rate of $d\dot{N}/dA\approx 10^{10}  (g^{3}_{aN})^{2} /({\rm cm^2s})$~\cite{CAST:2009klq}.
In contrast, axions produced via nuclear deexcitations, such as the 14.4 keV transition of $\iso{57}{Fe}$, offer significant advantages in environments where the temperature exceeds the excitation energy. At temperatures above $\mathord{\sim}14.4$ keV, a substantial fraction of $\iso{57}{Fe}$ nuclei populate their first excited state, greatly enhancing the axion production rate. This condition is naturally met in supergiant stars, whose core temperatures reach 20~keV or higher, making them particularly efficient axion sources. 

The axion emission rate per unit mass of stellar matter can be expressed as
\begin{equation}
\mathcal{\dot N}_a = \mathcal{N} \, \omega_1(T) \frac{1}{\tau_0} \frac{1}{1 + \alpha} \frac{\Gamma_a}{\Gamma\gamma},
\label{eq:emission_rate}
\end{equation}
where $\mathcal{N}$ is the ${ }^{57} \mathrm{Fe}$ number density per solar mass, $\omega_1$ the occupation number of the first excited state
\begin{equation}
\omega_1(T) = \frac{\left(2 J_1 + 1\right) e^{-E^* / T}}{\left(2 J_0 + 1\right) + \left(2 J_1 + 1\right) e^{-E^* / T}}\,,
\end{equation}
$\tau_0=141\,\text{ns}$ the lifetime of the excited state, $\alpha=8.56$ the internal conversion coefficient (see, e.g., Refs.~\cite{Avignone:2017ylv,DiLuzio:2021qct,OShea:2023gqn}) 
$E^* = 14.4$ keV the excitation energy, $T$ the local temperature, and $J_0$ and $J_1$ the angular momenta of the ground and excited states, respectively. 
Finally, the ratio $\Gamma_a / \Gamma_\gamma$ represents the branching fraction for axion emission relative to photon emission. In the case of $\iso{57}{Fe}$ deexcitation, 
\begin{equation}
\frac{\Gamma_a}{\Gamma_\gamma} = 2.32 \left( 0.16\, g_{ap} + 1.16\, g_{an} \right)^2\,.
\label{eq:ratio_Avignone}
\end{equation}
From the form of Eq.~\eqref{eq:ratio_Avignone}, 
it is convenient to define the effective nucleon coupling as~\cite{DiLuzio:2021qct}
\begin{align}
\label{eq:g_N_eff}
g_{aN}^{{\rm eff}}=0.16\, g_{ap} +1.16\, g_{an}\,,
\end{align}
which is the coupling combination that controls the axion emission rate in this transition.

For all practical purposes, the axion emission from the deexcitation of $^{57}\mathrm{Fe}$ appears as a spectral line for \mbox{\textsc{NuSTAR}}.
The line can be modeled by a Gaussian distribution centered at $E^* = 14.4~\mathrm{keV}$ and with a total width
$\sigma_{\mathrm{tot}}$. 
The largest contribution to the line width is the thermal spread~\cite{DiLuzio:2021qct},
\begin{equation}
\sigma_{\mathrm{th}}(T) = E^* \sqrt{T/m_{\rm Fe}},
\end{equation}
with $T$ the local temperature and $m_{\rm Fe}$ the mass of the $\iso{57}{Fe}$ nucleus. 
For the typical temperatures expected in the core of a supergiant star such as Betelgeuse during the post-MS evolutionary stages,
the thermal width is expected to be $\sigma_{\mathrm{th}}(T) \sim \mathcal{O}(10~\mathrm{eV})$, and can vary by a factor of a few over the course of the stellar evolution.
This is substantially narrower than the instrument resolution at the relevant energies. 
The width is further broadened by the rotation of the core.
To obtain a rough estimate of this contribution, we can start from Betelgeuse's equatorial velocity $v_{\mathrm{eq}}$. Reference~\cite{Kervella_18} reports $v_{\mathrm{eq}} \sin \theta \approx 5.47~\mathrm{km/s}$, 
where $\theta$ is the inclination angle between the rotation axis and the line of sight. This yields
$\sigma_{\mathrm{rot}} \approx 0.193~\mathrm{eV}$,
a value that remains negligible compared to thermal broadening, even when accounting for recent studies suggesting that core rotation in red-giant stars could be up to ten times faster than the observed equatorial velocity~\cite{Beck:2012}.

If future detectors were to be able to resolve this width, they could provide valuable insights into the core rotation—a parameter otherwise inaccessible to direct measurements.
Nonetheless, since current instrumentation cannot detect this effect, we disregard its impact in the subsequent analysis.

The expected differential axion flux at a distance $d$ from the source can be computed by integrating the Gaussian emission profile $\mathcal{G}(E, \sigma)$ over the stellar model,
\begin{align}
\frac{\mathrm{d} N_a\left(E\right)}{\mathrm{d} E} &=
\frac{1}{4 \pi d^2}
\int_0^{R_{\mathrm{star}}} \mathcal{N}_a(r)  \mathcal{G}(E, \sigma(r)) \rho(r) 4 \pi r^2 \dd r. \nonumber
\label{eqn:diff_flux}
\end{align}
For our numerical estimates, we used a stellar model for a supergiant of $20\,M_\odot$ and solar metallicity,
constructed using the MESA stellar evolution code~\cite{MESA_orig, MESA2013, MESA2015, MESA2018, MESA2019, MESA2023}.
Specifically, we extended the \texttt{20M\_pre\_ms\_to\_core\_collapse} MESA setup to include a significantly larger nuclear reaction network,
with a total of 206 isotopes, in order to properly account for all possible interactions that could affect the abundance of $^{57}\operatorname{Fe}$ throughout the evolution.
As the precise evolutionary stage of Betelgeuse is unknown, and $s$-process reactions can, in principle, affect the abundance of $^{57}\mathrm{Fe}$, we adopt a representative stellar model for our numerical results, corresponding to the onset of core carbon burning, after the $\iso{57}{Fe}$ abundance and therefore also the axion flux has stabilized.
In any case, the resulting bounds depend very weakly on the exact evolutionary stage.  
This contrasts with the case of continuum emission processes, such as the Primakoff and ABC mechanisms studied in Refs.~\cite{Xiao:2020pra,Xiao:2022rxk}, which exhibit a stronger sensitivity to variations in the core conditions—particularly the core temperature.

In our case, the core temperature during the post–He-burning evolution remains always above the excitation energy of the $^{57}\mathrm{Fe}$ nuclear level (cf. the left panel of Fig.~\ref{fig:mesa_radial}). 
As a result, the Boltzmann factor that governs the population of the excited state saturates, leading to a very mild dependence of the emission rate on temperature. 
Further details on the construction of our reference stellar model are provided in Appendix \ref{appx:mesa}.

The axions produced in the core of Betelgeuse can convert into hard x-ray photons in the Galactic magnetic field (GMF), providing an observational signature. 
The conversion probability in the regular magnetic field\footnote{We focus only on the regular component of the magnetic field in the main text, as it is by far the most relevant and leads to a conservative result. 
Our numerical study of the effects of the turbulent component of the GMF, following Ref. ~\cite{Carenza:2021alz}, gives a standard deviation of $\mathcal{O}(10\%)$ for the conversion probability, with no significant impact on the expected spectral shape of the signal. Details can be found in Appendix \ref{appx:gmf}.} is given by~\cite{Xiao:2022rxk},
\begin{equation}
P_{a \gamma}=8.7 \times 10^{-6} g_{11}^2\left(\frac{B_{\mathrm{T}}}{1 ~\mu \text{G}}\right)^2\left(\frac{d}{197 \text{ pc}}\right)^2 \frac{\sin ^2(q d)}{(q d)^2},
\end{equation}
where $g_{11} = (g_{a\gamma}/10^{11}~\text{GeV}^{-1})$ is a normalized axion-photon coupling, and $q$ is the momentum transfer term 
\begin{align}\label{Eq:qd}
q d &\simeq 
\left[77\left(\frac{m_a}{10^{-10} \text{ eV}}\right)^2 
- 0.14\left(\frac{n_e}{0.013 \text{ cm}^{-3}}\right) \right] \notag \\
&\quad \times \left(\frac{d}{197 \text{ pc}}\right) 
\left(\frac{E}{1 \text{ keV}}\right)^{-1}
\end{align}
Here, $m_a$ is the axion mass, $n_e$ the electron number density \cite{cordes2003ne2001inewmodelgalactic} and $E$ the axion energy. 
Throughout this study, we assume a constant transverse magnetic field of $B_{\mathrm{T}}=1.4~\mu$G~\cite{XuHan_2019}, 
though we present also the results for the more optimistic scenario of  $B_{\mathrm{T}}=3.0~\mu$G~\cite{Harvey_Smith_2011}. 
The exact value of the magnetic field remains the main source of uncertainty in this work, as the final signal scales $\mathord{\sim}B_{\mathrm{T}}^2$, yielding $P_{a \gamma}=(1.5 - 9.2)\times 10^{-5}$ for the parameters used in this work. 
This range includes the effects of a turbulent component of the magnetic field $B^2_{\mathrm{rms, turb.}}\approx (1 \ \mu G)^2$ with a correlation length of $\mathcal{O}(10 \ \mathrm{pc})$, see Appendix \ref{appx:gmf} for details.
Finally, notice that our results are conservative as we are neglecting the contribution of the local magnetic field of Betelgeuse.
\footnote{Although the existence of the magnetic field of Betelgeuse is established (see, e.g., Ref.~\cite{Auriere:2010cb}), both its strength and structure remain highly uncertain, with indications of temporal variability and complex spatial morphology. 
Including this component would mainly affect the high-mass region of the ALP parameter space (see Ref.~\cite{Manzari:2024jns}), while our conclusions at low masses are dominated by the well-characterized Galactic field. We therefore adopted the conservative choice of excluding the local field.}

\begin{figure*}[!t]
    \centering
    \begin{subfigure}[b]{0.49\textwidth}
        \centering
        \includegraphics[width=\textwidth]{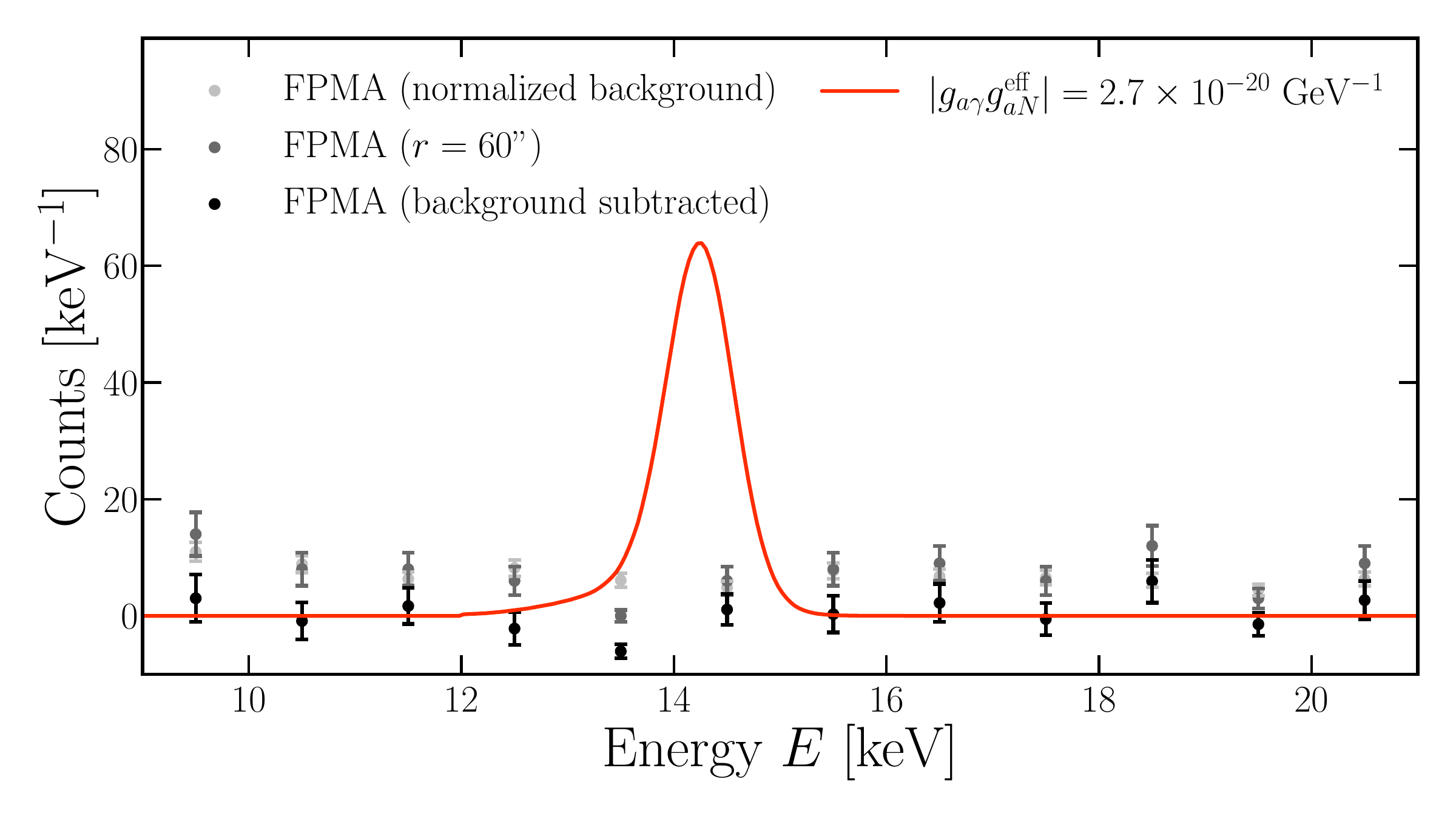}
        \label{fig:FPMA_spectrum}
    \end{subfigure}
    \begin{subfigure}[b]{0.49\textwidth}
        \centering
        \includegraphics[width=\textwidth]{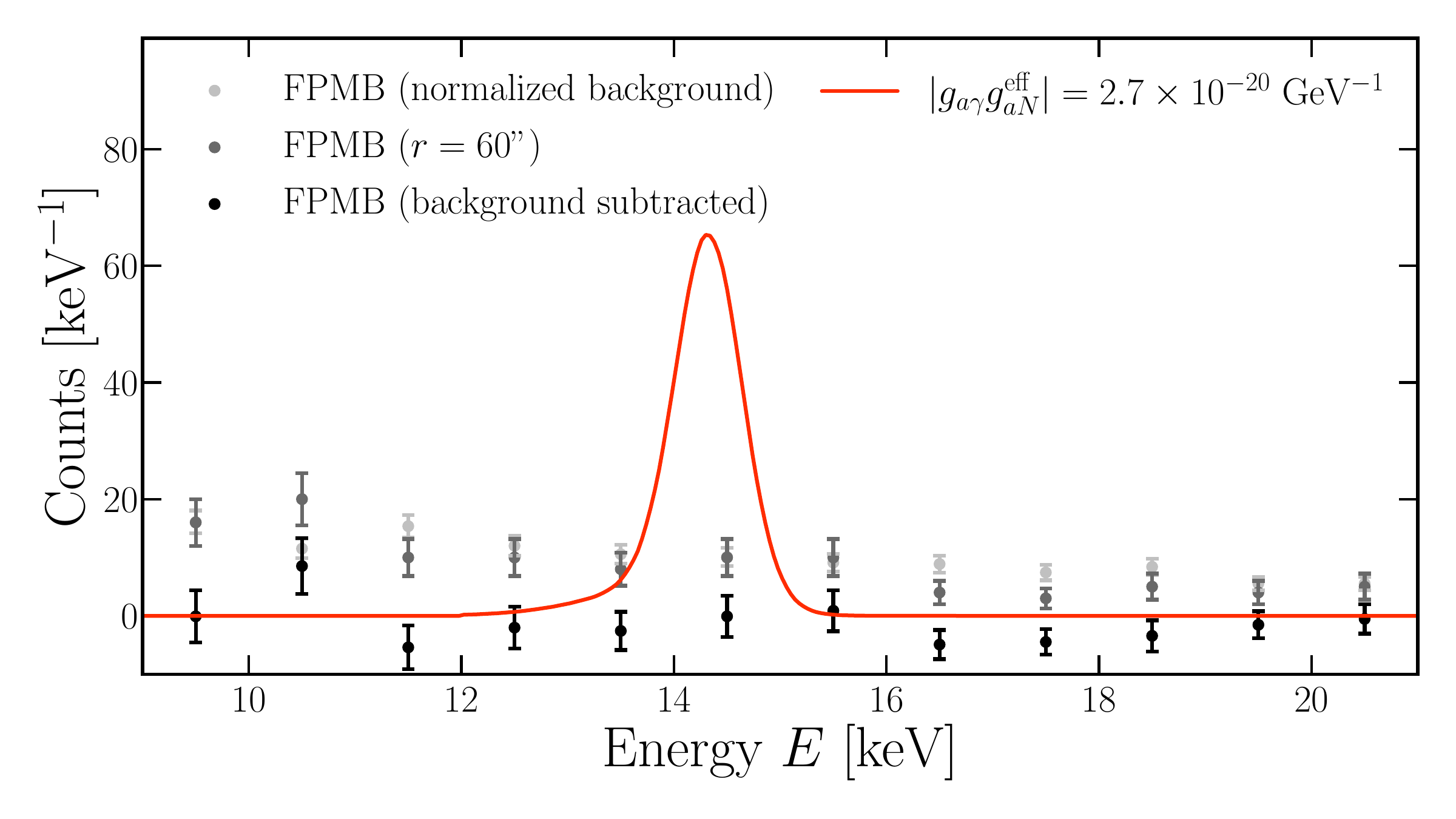}
        \label{fig:FPMB_spectrum}
    \end{subfigure}
    \caption{Expected axion signal in the two modules FPMA (\textit{left}) and FPMB (\textit{right}). Shown is the signal for an effectively massless axion ($m_a \lesssim 10^{-10}$ eV) assuming the maximally allowed value for the combination of the couplings (at 95\% CL) $|g_{a\gamma}g_{aN}^{\mathrm{eff}}| = 2.7\times 10^{-20}$ GeV$^{-1}$. The spectra are binned in 1 keV increments and shown in the 9–21 keV range for visualization purposes, while the analysis was conducted on the full energy range using unbinned data.
    }
    \label{fig:spectra}
\end{figure*}

\section{NuSTAR data analysis and results}
For our work, we have analyzed the only available \textsc{NuSTAR} observation of Betelgeuse, carried out on August 23, 2019 (ObsID~30501012002). 
Below, we present our data analysis and the main results obtained. 
Further details can be found in Appendix \ref{appx:NUSTAR}.
The \textsc{NuSTAR} Observatory~\cite{NuSTAR:2013yza}, the first high-energy focusing x-ray telescope in space, operates within the $3$--$79$~keV energy range, which makes it well-suited for investigating the expected axion-induced $^{57}\mathrm{Fe}$ deexcitation signal at $14.4$~keV. \textsc{NuSTAR} consists of two co-aligned telescope modules, each equipped with independent optics and focal-plane module (FPM) Cadmium-Zinc-Tellurium (CdZnTe) detectors, designated as FPMA and FPMB. Each module provides a field of view of approximately $13' \times 13'$ and achieves an angular resolution of $58''$ half-power diameter for point sources near the optical axis. These grazing-incidence optics efficiently direct x-rays onto the focal-plane detectors, ensuring high sensitivity and resolution for the purposes of our study.

Data from both detectors were processed using the \texttt{HEASoft} software (version~6.34)~\cite{ascl:1408.004} together with the \texttt{NuSTARDAS} package (version~2.1.4). The total cleaned exposures were found to be 49.16~ks for FPMA and 48.83~ks for FPMB.
A circular source region with a radius of 60$''$ was selected, centered on the J2000 coordinates of Betelgeuse: 
$ \text{RA} = 88^\circ\,47'\,34.8'', \quad \text{Dec} = +07^\circ\,24'\,25.4'' $. 
Following the approach used in \cite{Xiao:2020pra, Xiao:2022rxk}, the background spectrum is obtained from a polygonal region separated $120''$ from the center of the source and located in the same CdZnTe chip that encompasses the source (see Appendix \ref{appx:background}). 
The x-ray spectra observed by FPMA and FPMB in both the source and background regions are displayed in Figure \ref{fig:spectra}.

The expected counts coming from the initial axion signal at \textsc{NuSTAR} FPMs, for a given set of nuisance parameters 
\(\boldsymbol{\theta} = \{g_{a\gamma}, g_{aN}, m_a\}\), are computed by forward modeling the photon signal from axion decay through the instrument response files and exposure times of each module (see Appendix \ref{appx:NUSTAR} for a detailed explanation).In Fig. \ref{fig:spectra}, the expected signal is shown for both detector modules, FPMA and FPMB.

\begin{figure}[t]
    \centering
    \includegraphics[width=1\linewidth]{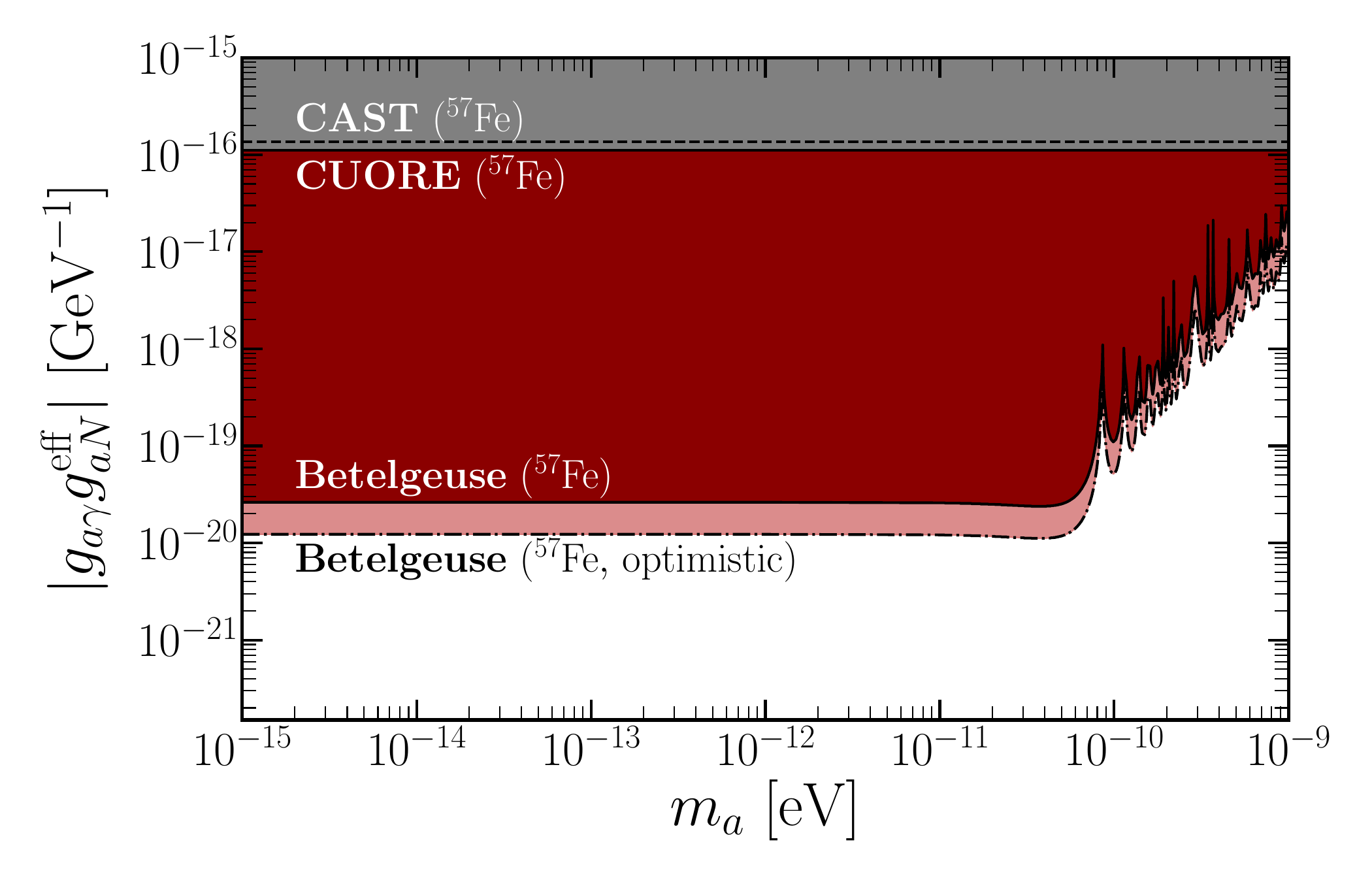}
    \caption{Exclusion limits from the Bayesian unbinned likelihood analysis at 95\% CL for massive axions. Shown are bounds from searches for the $14.4$ keV line of $^{57}$Fe by CAST \cite{CAST:2009jdc} and CUORE \cite{Li:2015tyq}.}
    \label{fig:results_exclusion}
\end{figure}

For each module $i=\{\mathrm{A}, \mathrm{B}\}$, the total observed counts \(N_{\mathrm{obs}, i}\) and total background counts \(N_{\mathrm{bkg}, i}\) are then combined in an unbinned likelihood. In order to obtain an upper limit on the coupling product, we introduce the unbinned likelihood function 

\begin{equation}
\mathcal{L}=\prod_{i=\{\mathrm{A},\mathrm{B}\}}
\mathcal{L}_i \times \prod_{i=\{\mathrm{A},\mathrm{B}\}} \mathcal{G}\left(\delta_{\mathrm{bkg}}^i\, \sigma_{\mathrm{bkg}}^i\right),
\end{equation}

and the individual likelihood, e.g. \(\mathcal{L}_i\), is computed as
\begin{equation}
\begin{split}
& \mathcal{L}_i 
= \mathrm{Poisson}\Bigl(N_{\mathrm{obs}, i} \mid N_{\mathrm{exp}, i}\Bigr) \\[1mm]
&\times \prod_{j=1}^{N_{\mathrm{obs}, i}}
  \left[
      \frac{N_{\mathrm{ax}, i} P_{\mathrm{ax}}(E'_j)}{N_{\mathrm{exp}, i}}
      + \frac{N_{\mathrm{bkg}, i}\,(1 + \delta_{\mathrm{bkg}}^i)\,P_{\mathrm{bkg}}(E'_j)}{N_{\mathrm{exp}, i}}
  \right]
\end{split}
\end{equation}
\noindent
where the first term is the usual Poisson distribution,
\begin{align}
 \operatorname{Poisson}\left(N_{\mathrm{obs}, i}\mid N_{\mathrm{exp}, i}\right) &= \frac{N_{\mathrm{exp}, i}^{\,N_{\mathrm{obs}, i}} e^{-N_{\mathrm{exp}, i}}}{N_{\mathrm{obs}, i}!}.
\end{align}
Here, the expected number of events for an axion signal is given by
$
N_{\mathrm{exp}, i} = N_{\mathrm{ax}, i} + N_{\mathrm{bkg}, i}\,(1+\delta_{\mathrm{bkg}}^i),
$
Here, \(N_{\mathrm{ax}, i}\) represents the number of axion signal events calculated via forward modeling of the photon signal resulting from axion decay. The energy-dependent axion signal probability density function, \(P_{\mathrm{ax}}(E_j)\), is defined for a specified axion mass \(m_a\), photon coupling \(g_{a\gamma}\), and transverse magnetic field \(B_T\). In the corresponding product, it is evaluated at the energy \(E_j\) that corresponds to the \(j\)th count. For both FPMA and FPMB, the fractional systematic uncertainty of the background, \(\sigma_{\mathrm{bkg}, i}\), is set to 10\%, with independent Gaussian fluctuations \cite{Xiao:2022rxk}. Since the background normalization uncertainty is treated as a nuisance parameter, \(\delta_{\mathrm{bkg}}^i\) is marginalized over the interval 
$
\delta_{\mathrm{bkg}}^i \in [-1, +\infty).
$
With a flat prior on the parameter vector \(\theta\), Bayes’ theorem yields the posterior probability
\begin{equation}
    P(\theta)=\frac{\mathcal{L}(\theta)}{\int \mathcal{L}(\theta')\,\dd\theta'},
\end{equation}
and the 95\% credibility upper limit on the coupling product is determined by solving
\begin{equation}
    \int_0^{(g_{a\gamma}^2\,g_{aN}^2)_{95\%}} P\Bigl(g_{a\gamma}^2\,g_{aN}^2\Bigr)\, \dd\Bigl(g_{a\gamma}^2\,g_{aN}^2\Bigr)=0.95.
\end{equation}
For the case of a massless axion, this yields 
\begin{equation}
    |g_{a\gamma}g_{aN}^{\mathrm{eff}}| \leq (1.2-2.7)\times10^{-20}~\text{GeV}^{-1},
\end{equation}
at 95\% CL. This result is valid for masses up to $m_a\lesssim 10^{-10}$ eV, when the axion-photon conversion loses coherence. The results for the massive case are displayed in Fig. \ref{fig:results_exclusion}. 
\section{Discussion}
In this work, we have presented the first search for axion-nucleon interactions via nuclear deexcitations in a supergiant star, using \textsc{NuSTAR} observations of Betelgeuse. 
Exploiting the elevated core temperature of this evolved star, we have shown that the 14.4~keV transition of $\iso{57}{Fe}$ offers a promising probe of the axion-induced monochromatic x-ray signature. 
Our analysis, based on a forward modeled Bayesian likelihood approach, reveals no significant excess at the expected energy but leads to the strongest direct bounds on the coupling product $|g_{a\gamma}g_{aN}^{\mathrm{eff}}|$ for $m_a \lesssim 10^{-10}$~eV, reaching values below $2.7\times10^{-20}~\text{GeV}^{-1}$.

These results establish supergiant stars as a powerful new laboratory for axion searches, complementing solar-based and terrestrial efforts. 
The expected stability of the $\iso{57}{Fe}$-induced signal across stellar evolutionary phases and the narrow, line-like nature of the x-ray signature make this channel especially attractive. 
Unlike continuum processes such as Primakoff or Compton production~\cite{Xiao:2020pra, Xiao:2022rxk}, which scale strongly with temperature and stellar evolutionary stage, the axion emission from $\iso{57}{Fe}$ saturates rapidly once the temperature exceeds 14.4~keV. 
As a result, this process constitutes a more robust probe across different evolutionary phases of the star. 
The main source of uncertainty is thus reduced to the strength of the Galactic magnetic field, while the specific evolutionary stage plays a less significant role than in the case of continuum processes.

To complete our analysis, we now relate our bound to the very strong astrophysical bounds derived from the cooling of supernova (SN) 1987A~\cite{Carenza:2019pxu,Carenza:2020cis}, as well as from observations on the cooling of neutron stars~\cite{Buschmann:2021juv}. 
Both arguments provide the approximate bound $g_{aN}\lesssim 10^{-9}$, where with $g_{aN}$ we indicate a generic axion-nucleon coupling~\cite{Carenza:2024ehj}. 
Let us take the SN bound in Ref.~\cite{Carenza:2019pxu} as a our concrete example.
Using the methods discussed in Refs.~\cite{DiLuzio:2021qct,Lucente:2022esm}, we see that $g_{aN}^{\mathrm{SN}}=10^{-9}$ corresponds to our $g_{aN}^{\mathrm{eff}}=0.9\times 10^{-9}$. 
If we  fix the axion-nucleon coupling to this value and translate our result into a bound on the photon coupling, we find $g_{a\gamma} \lesssim 1.1 \times 10^{-11}\mathrm{GeV}^{-1}$, which is slightly stronger than the current CAST limit and lies within the region targeted by future helioscope experiments~\cite{IAXO:2019mpb,IAXO:2020wwp,IAXO:2024wss}.
We finally notice that the region explored in this work is in tension with the bound derived from the nonobservation of a gamma ray flux associated to the SN 1987A event, by the Solar Maximum Mission~\cite{Payez:2014xsa,Hoof:2022xbe}.
In particular, Ref.~\cite{Calore:2020tjw} provides the bound $g_{a \gamma}<3.4 \times 10^{-15}\, \mathrm{GeV}^{-1}$ for $m_a\lesssim 0.1$ neV, in the case of $g_{aN}^{\mathrm{SN}}=10^{-9}$, with $g_{aN}^{\mathrm{SN}}$ the effective coupling entering in the axion bremsstrahlung production in SN~\cite{Carenza:2019pxu,Lella:2022uwi,Lella:2023bfb}.
Using 
$g_{aN}^{\mathrm{eff}}=0.9\times 10^{-9}$, we see that  
the bound extracted in Ref.~\cite{Calore:2020tjw} translates to $|g_{a\gamma}g_{aN}^{\mathrm{eff}}|\lesssim 3\times 10^{-24}\, \mathrm{GeV}^{-1}$ for $m_a \lesssim 10^{-10}$~eV.
We have not reported this bound in our figures as it was not a direct measurement.~\footnote{In fact, as explicitly discussed in Ref.~\cite{Payez:2014xsa}, the telescope was not even pointing at the SN location at the time of the explosion.}

Future observations of additional nearby supergiants, together with advancements in x-ray instrumentation and improved modeling of the Galactic magnetic field, hold strong potential to further tighten current constraints—or even lead to the first detection of an axion signal from stellar nuclear transitions.
Moreover, as first shown in Refs.~\cite{Ning:2024eky,Ning:2025tit, Candon:2024eah}, entire galaxies can act as competitive axion sources. In particular, starburst galaxies such as M82, with their high abundance of supergiant stars, offer promising laboratories to test the axion-nucleon coupling via $\iso{57}{Fe}$ decay.
Preliminary studies indicate that the analysis of \textsc{NuSTAR} observations of M82 could yield an even more stringent bound on $|g_{a\gamma}g_{aN}^{\mathrm{eff}}|$ than the one derived here.
A detailed study along these lines is left for future work.

Data analysis was performed using the \textsc{NuSTAR} Data Analysis Software, which was jointly developed by the ASI Science Data Center in Italy and the California Institute of Technology in the United States.

\begin{acknowledgements} 
We warmly thank Diego Guadagnoli, Guiseppe Lucente and Edoardo Vitagliano for helpful comments on the draft. F.\,R.\,C. and M.\,K. acknowledge insightful discussions with Igor Irastorza about statistics.
This article is based upon work from COST Action COSMIC WISPers (CA21106). F.\,R.\,C. is supported by the Universidad de Zaragoza under the “Programa Investigo” (Programa Investigo-095-28), as part of the Plan de Recuperación, Transformación y Resiliencia, funded by the European Union-NextGenerationEU.
The work of M.\,G. and M.\,K. is supported by the grant PGC2022-126078NB-C21 funded by MCIN/AEI/ 10.13039/501100011033 and “ERDF A way of making Europe”. M.\,G. and M.\,K. further acknowledge the grants DGA-FSE 2023-E21-23R and DGA-FSE 2020-E21-17R, respectively. Both are funded by the Aragon Government and the European Union – Next Generation EU Recovery and Resilience program on `Astrofísica y Física de Altas Energías' CEFCA-CAPA-ITAINNOVA.
M.\,G. acknowledges funding from the European Union’s Horizon 2020 research and innovation programme under the European Research Council (ERC) grant agreement ERC-2017-AdG788781 (IAXO+).
M.\,K. is further supported by the Government of Aragón, Spain, with a PhD fellowship as specified in ORDEN CUS/702/2022.\\
\end{acknowledgements} 

\section*{Data Availability}

The data that support the findings of this article are
openly available \cite{NuSTAR_ObsID30501012002}.

\appendix
\section{MESA Simulation Details}
\label{appx:mesa}
Our numerical stellar model, representing a $20M_\odot$ star with metallicity $Z=0.02$,  
was produced with MESA \texttt{v.24.08.1}, a publicly available one-dimensional stellar evolution code, which solves the equations of stellar structure and returns stellar profiles throughout the stellar evolutionary stages \cite{MESA_orig, MESA2013, MESA2015, MESA2018, MESA2019, MESA2023}.
In order to account for the several processes which might impact the abundance of $^{57}\operatorname{Fe}$, we extended the nuclear  
reaction network significantly, including a total of 206 isotopes.~\footnote{This network is provided in the MESA framework as the file ``\texttt{mesa\_206.net}'', and is primarily used in presupernova models.} 
Our final models are quite similar to the ones used in Refs.~\cite{Xiao:2020pra,Xiao:2022rxk}, which we used as a reference for the stellar properties, except for the much larger nuclear network. 

\begin{figure}[h]
    \centering
    \includegraphics[width=\linewidth]{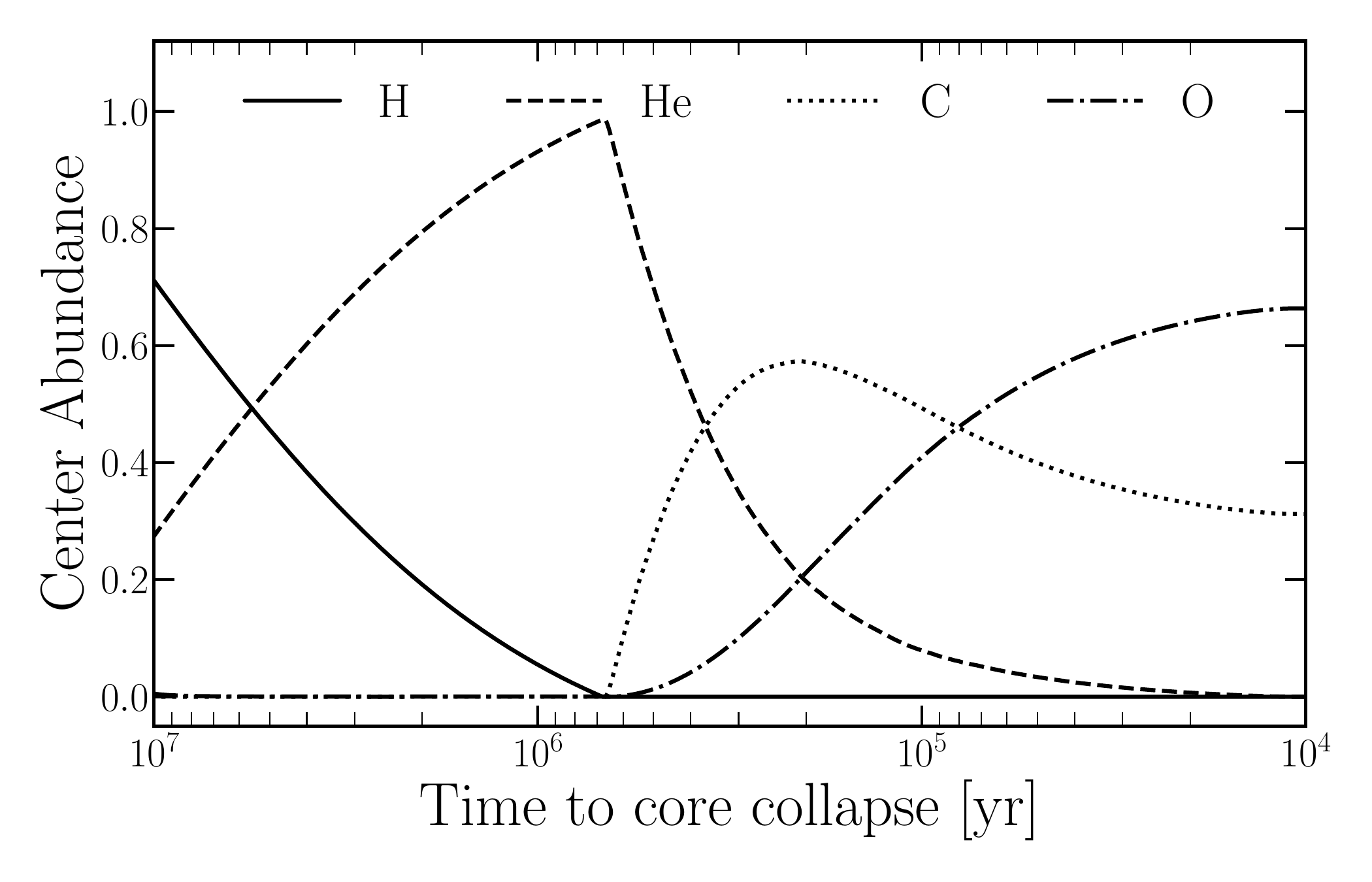}
    \caption{Central abundances of hydrogen, helium, carbon, and oxygen for the supergiant star as a function of time. Tracking the abundances allows one to identify the key stages of the stellar evolution, like the onset of helium and carbon burning in the stellar core.}
    \label{fig:mesa1}
\end{figure}
\begin{figure*}[tbp]
    \centering
    \begin{subfigure}[b]{0.49\textwidth}
        \centering
        \includegraphics[width=\textwidth]{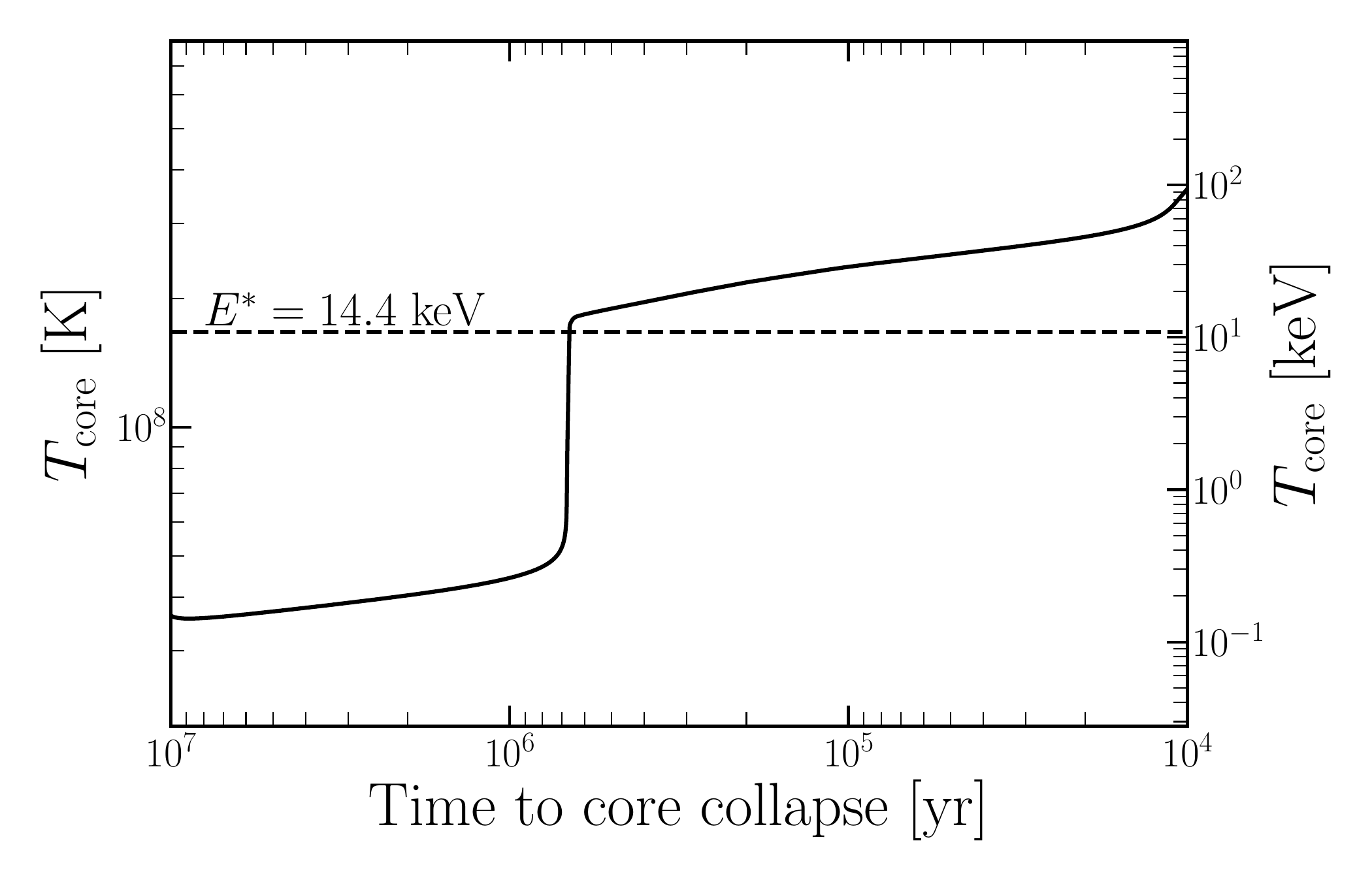}
        \label{fig:T_core}
    \end{subfigure}
    \hfill
     \begin{subfigure}[b]{0.49\textwidth}
        \centering
        \includegraphics[width=\textwidth]{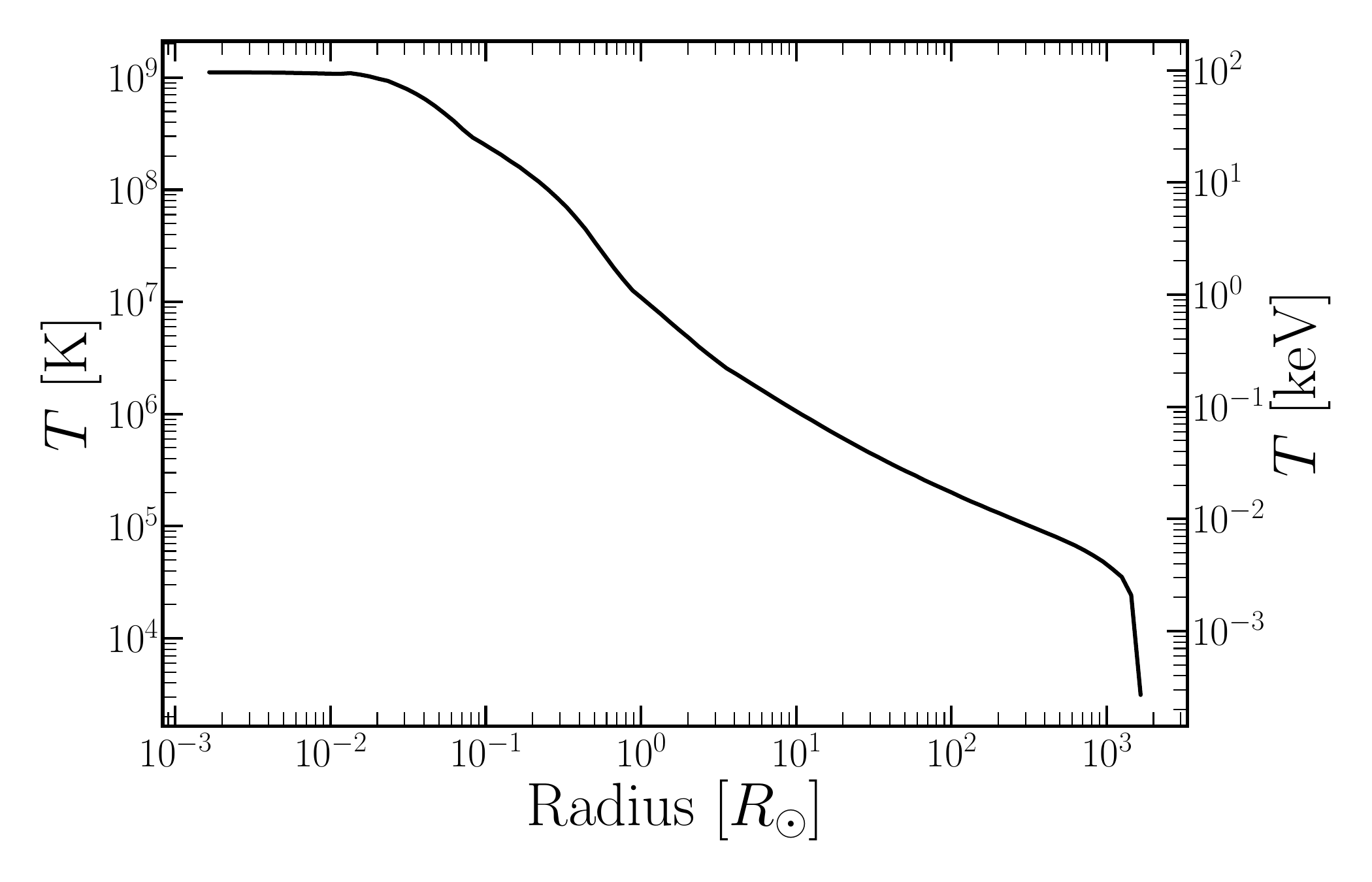}
        \label{fig:rho_core}
    \end{subfigure}
    \caption{\textit{Left}: stellar core temperature as a function of time to core collapse for our simulation. \textit{Right}: radial temperature profile of our benchmark $20M_\odot$ stellar model, taken at a time when the carbon burning phase is about to finish.}
    \label{fig:mesa_radial}
\end{figure*}

In Fig.~\ref{fig:mesa1}, we show the time evolution of the central abundances in our $20\,M_\odot$ supergiant star model.
The ignition of helium in the core, occurring about 0.7 Myr before core collapse (CC), is evident as the helium abundance begins to decrease while carbon and oxygen abundances increase.
This marks the transition to the red giant phase.
As it is well known, throughout the stellar evolution, heavier and heavier elements are ignited in the core, up to the formation of an inert iron core which, being unable to release energy through fusion, triggers gravitational collapse and signals the end of its life. 
The core temperature of the star increases steadily during the post–core-helium-burning phase, as shown in the right panel of Fig.~\ref{fig:mesa_radial}, and remains consistently above the $^{57}\mathrm{Fe}$ excitation energy of 14.4 keV, marked as a dashed line.
As discussed in the text, the temperature dependence of the deexcitation rate—which rises steeply for $T < E^*$—becomes very weak above the excitation temperature, as the Boltzmann exponential saturates.
This represents a substantial difference between our process and other continuous mechanisms, such as the ABC and Primakoff processes studied in Refs.\cite{Xiao:2020pra,Xiao:2022rxk}.

In the right panel of Fig.~\ref{fig:mesa_radial}, we show the radial temperature profile at a single time step for our $20\,M_\odot$ supergiant star. Specifically, the selected time step is chosen to avoid major evolutionary transition periods, when the axion flux has stabilized to its post-He ignition value.

Apart from temperature, the abundance of $\iso{57}{Fe}$ is another key parameter influencing the axion deexcitation rate. We find that after the star enters the red giant regime, the $\iso{57}{Fe}$ abundance remains approximately constant, with a notable exception: Right at the ignition of $\iso{}{He}$ burning, we observe an increased neutron density which sources $\iso{57}{Fe}$ via neutron capture (i.\,e. $s$ processes such as $\iso{56}{Fe}+n \rightarrow \iso{57}{Fe}+\gamma$)  or neutron-proton exchange (e.g. $\iso{57}{Co}$ + $n \, \rightarrow$ \, $\iso{57}{Fe}$ + $p$). 
Subsequently, the increase in the abundance is countered by reactions such as $\iso{57}{Fe} + n \, \rightarrow \iso{58}{Fe} + \gamma$, leading to a relaxation of the abundance to roughly the value from right at the beginning of $\iso{}{He}$ burning. This value is stable from the onset of carbon burning until just before core collapse, which justifies the choice of any profile from early $\iso{}{He}$ burning or after the onset of $\iso{}{C}$ burning for a conservative estimate of the axion flux.  

\section{Impact of the Turbulent component of the Galactic Magnetic Field}\label{appx:gmf}

\begin{figure*}[tbp]
    \centering
    \includegraphics[width=.8\linewidth]{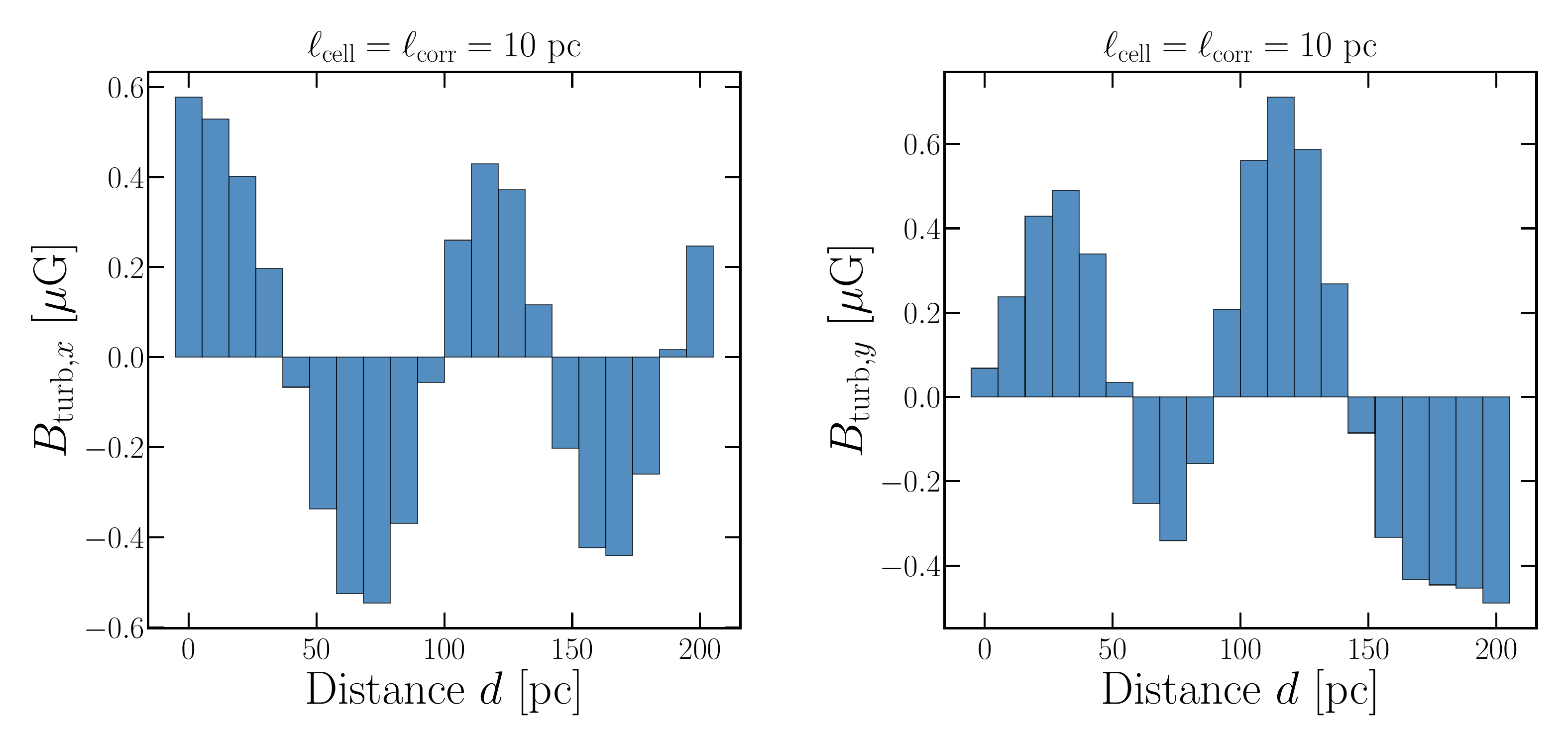}
    \caption{Visualization of an example realization of a turbulent component of the GMF with $B^2_{\mathrm{rms, turb.}} = (1.0\ \mu G)^2$ and $\ell_{\mathrm{corr}} = 10\ \mathrm{pc}$, as computed with the \texttt{gammaALPs} code \cite{Meyer:2021pbp}.}
    \label{fig:B_turb}
\end{figure*}

The results shown in the main text (including in the figures) consider only the regular component of the magnetic field. 
However, as extensively discussed in Ref.~\cite{Carenza:2021alz}, radio observations of synchrotron emission and polarization reveal the presence of an additional small-scale component of the magnetic field, superimposed on the regular one. 
This turbulent (or random) component has a comparable field strength of $\mathcal{O}(1)\,\mu\mathrm{G}$ but a much shorter correlation length of $\mathcal{O}(10$–$100)\,\mathrm{pc}$ than the regular field.
The analysis in Ref.~\cite{Carenza:2021alz} showed that this turbulent component can have a non-negligible impact on the ALP–photon oscillation probability, making a dedicated study necessary for each specific case.
Employing the methodology of Ref.~\cite{Carenza:2021alz}, we studied the impact of additional different \textit{random} Gaussian fields on our results for the characteristic $14.4\ \mathrm{keV}$ line of $^{57}\mathrm{Fe}$. One such realization of a Gaussian random field with $B^2_{\mathrm{rms, turb.}}= (1\ \mu G)^2$ and $\ell_{\mathrm{corr}}=10\ \mathrm{pc}$ as computed using the \texttt{gammaALPs} code~\cite{Meyer:2021pbp} is displayed in Fig.~\ref{fig:B_turb}.
Here, the power spectrum of the turbulent field follows a power law, $M(k)\sim k^q$, with spectral index $q= -11/3$, known as Kolmogorov spectrum, and is normalized with respect to $B^2_{\mathrm{rms, turb.}}$. 
The Fourier transform of the correlation function of the transversal component of the $B$ field, $B_T$, which is the relevant one for the conversion process, is given by 
\begin{equation}
\tilde{\epsilon}_{\perp}(k)=\frac{\pi B^2_{\mathrm{rms, turb.}}}{4} F_q\left(k ; k_L, k_H\right),
\end{equation}
where the functional form of $F_q\left(k ; k_L, k_H\right)$ is explicitly derived in the Appendix of Ref. \cite{Meyer:2014epa}. 

\begin{figure*}[tbp]
     \centering
     \begin{subfigure}[b]{0.49\textwidth}
         \centering
         \includegraphics[width=\textwidth]{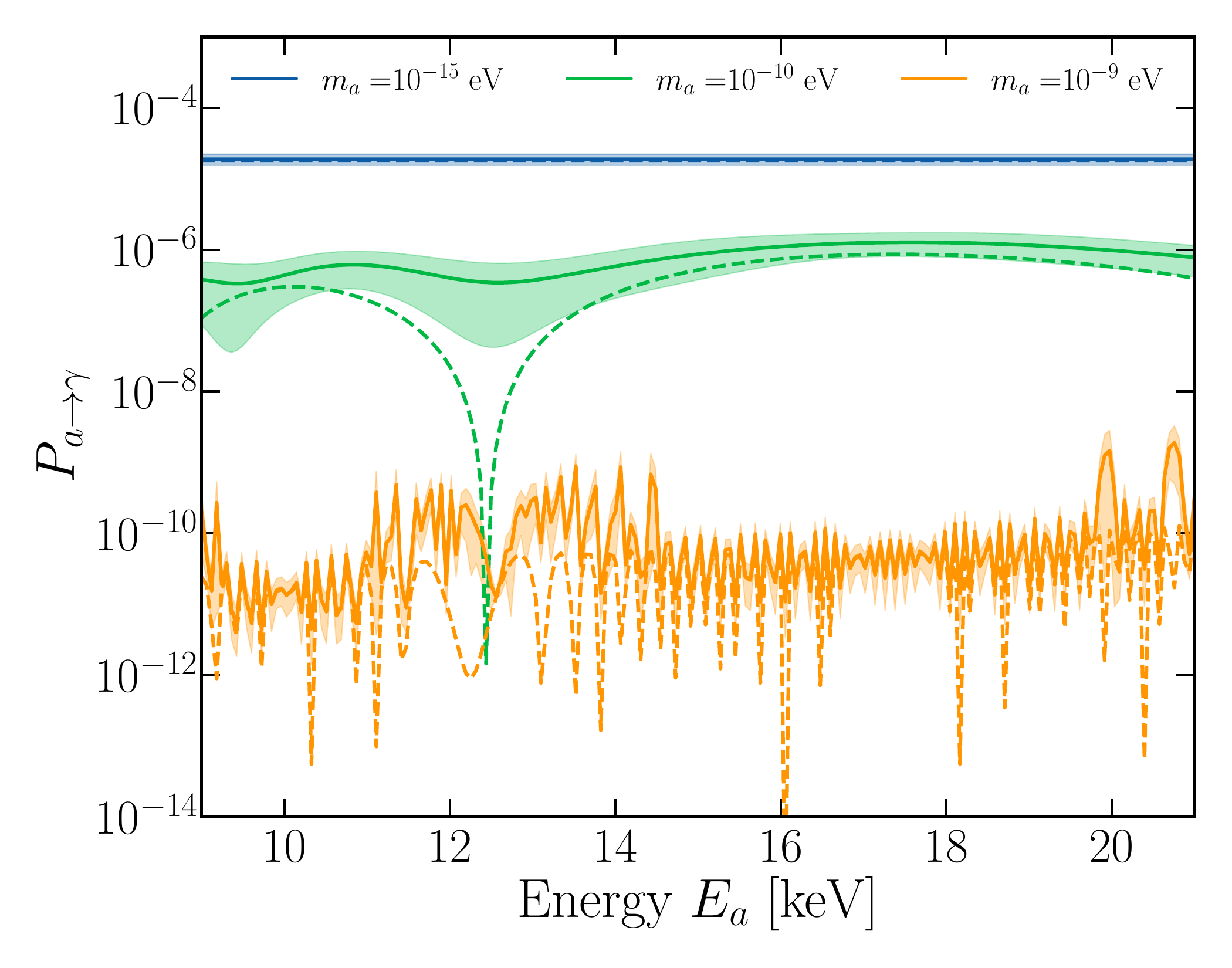}
         \label{fig:P_spec}
     \end{subfigure}
     \hfill
      \begin{subfigure}[b]{.49\textwidth}
         \centering
         \includegraphics[width=\textwidth]{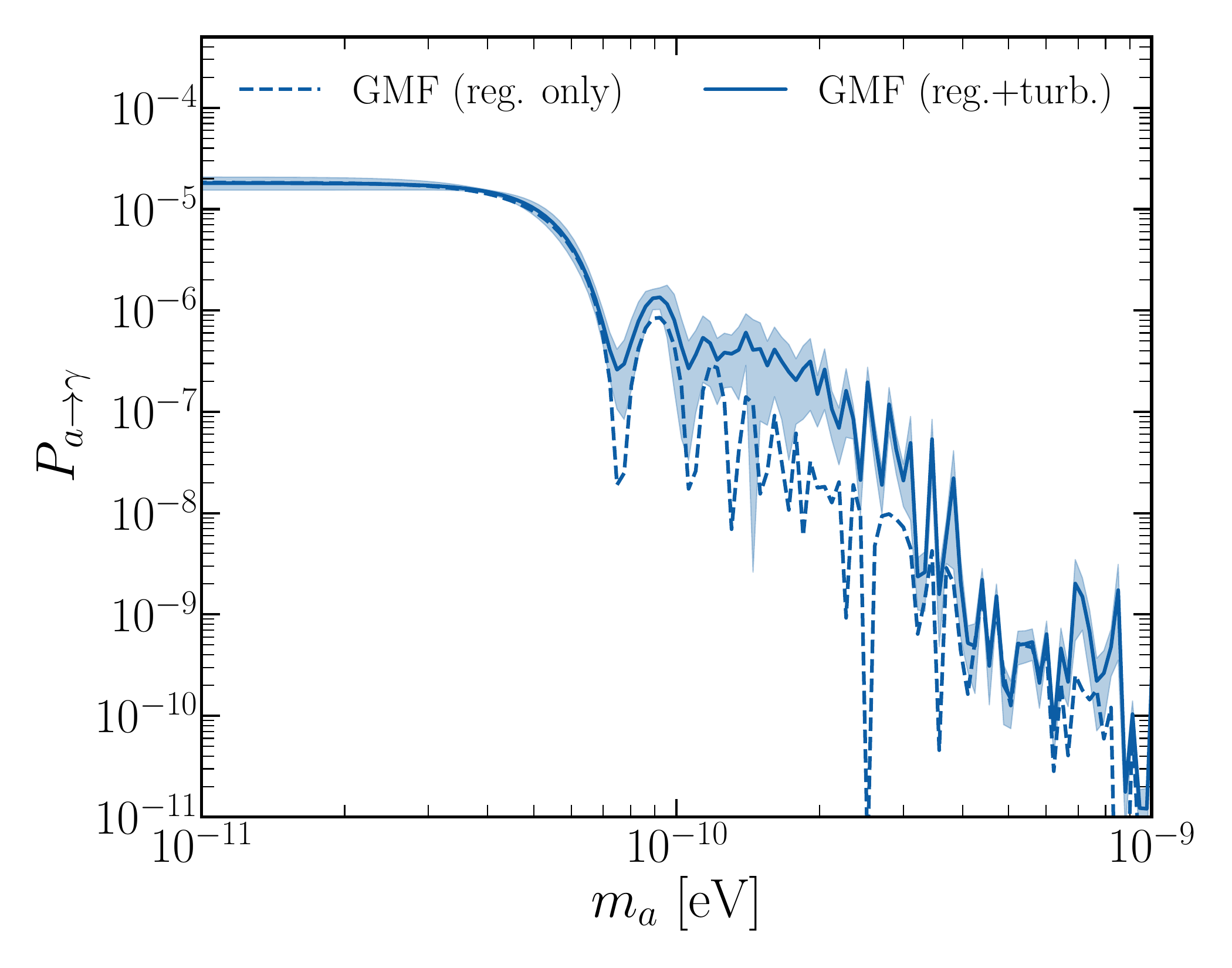}
         \label{fig:P_mass}
     \end{subfigure}
     \caption{\textit{Left}: axion-photon conversion spectrum $P_{a\to\gamma}(E_a)$ in the relevant energy range for the expected signal at \textsc{NuSTAR}, cf. Fig. \ref{fig:spectra}, with $g_{a\gamma}=10^{-11}\ \mathrm{GeV}^{-1}$ and $B_{T, \mathrm{reg.}} = 1.4 \ \mu G$. \textit{Right}: mass-dependent axion-photon conversion probability $P_{a\to\gamma}$ for a fixed axion energy of $E_a=14.4\ \mathrm{keV}$, $g_{a\gamma}=10^{-11}\ \mathrm{GeV}^{-1}$ and $B_{T, \mathrm{reg.}} = 1.4 \ \mu G$. In both plots we show the regular-only GMF case, as used for the main part of the work, as a dashed line and the mean and standard deviation of a series of 20 random realizations including a turbulent component of the GMF of $B^2_{\mathrm{rms, turb.}} = (1.0\ \mu G)^2$ similar to the example shown in Fig. \ref{fig:B_turb}, as a solid line with a shaded band.}
     \label{fig:conversion_study}
 \end{figure*}

We estimated the effects of the turbulent component of the magnetic field by simulating 20 realizations of \textit{random} Gaussian fields and comparing the resulting mean and standard deviation to the result, which includes only the regular field.
Figure~\ref{fig:conversion_study} shows our findings, with the dashed lines corresponding to the regular magnetic field alone and the colored bands representing the impact of the random component.
As expected, the effects of the turbulent magnetic field are practically limited to the larger mass case. 
In fact, while the ALP-photon oscillations are coherent, the oscillation probability scales approximately as $(LB)^2$, with $L$ the correlation length and B the strength of the magnetic field component, and thus the regular component dominates. 
However, at higher ALP mass, when the oscillations induced by the regular magnetic field lose coherence, the effect of the turbulent field becomes more relevant, as the shorter correlation length raises the threshold mass at which coherence is lost [cf. Eq.~\eqref{Eq:qd}].

\section{Data Reduction and NuSTAR Analysis}\label{appx:NUSTAR}
\begin{figure*}[!t]
    \centering
    
    \begin{subfigure}[b]{0.49\textwidth}
        \centering
        \includegraphics[width=\textwidth]{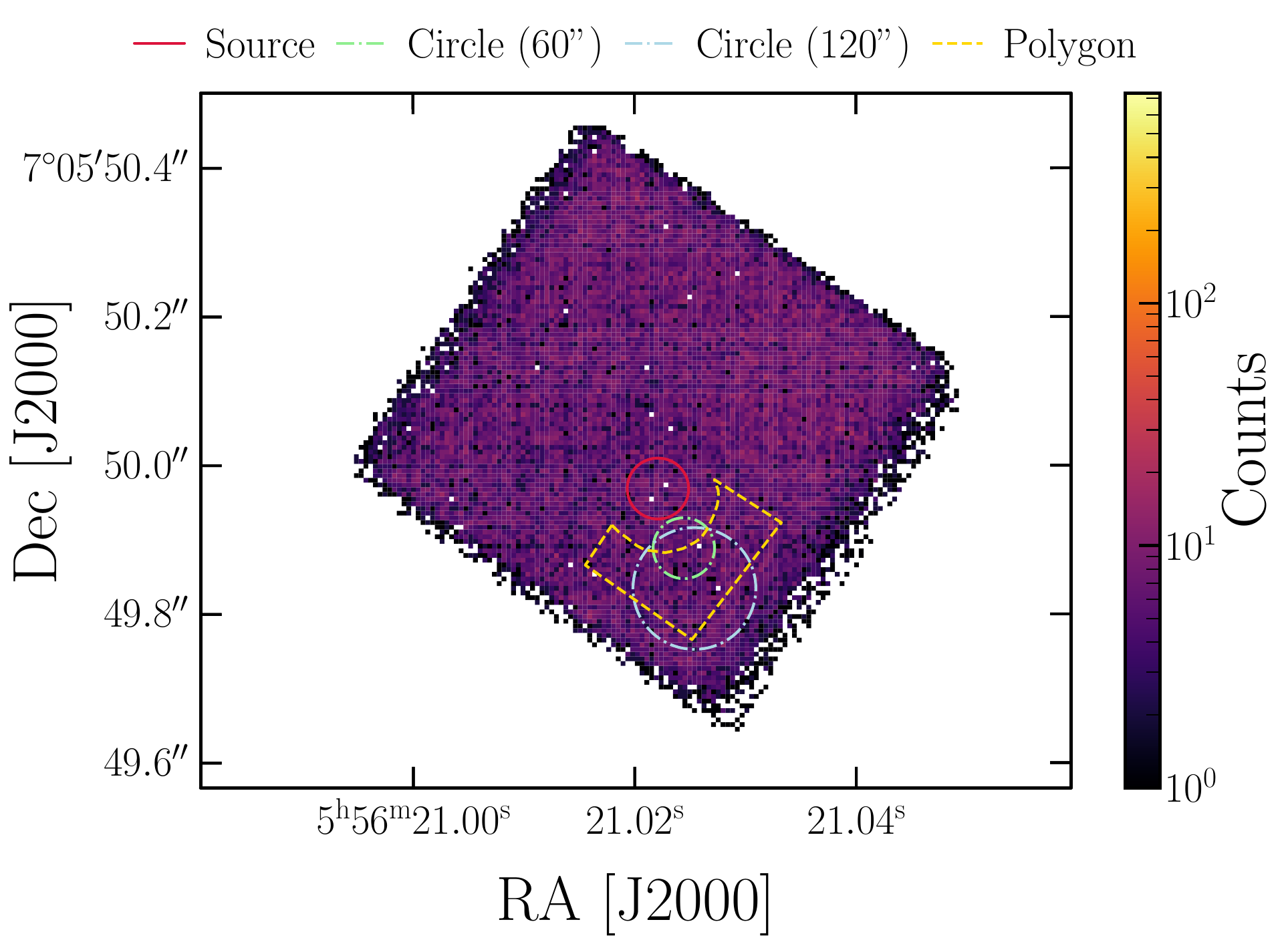}
        \label{fig:FOV_A}
    \end{subfigure}
    \begin{subfigure}[b]{0.49\textwidth}
\includegraphics[width=\textwidth]{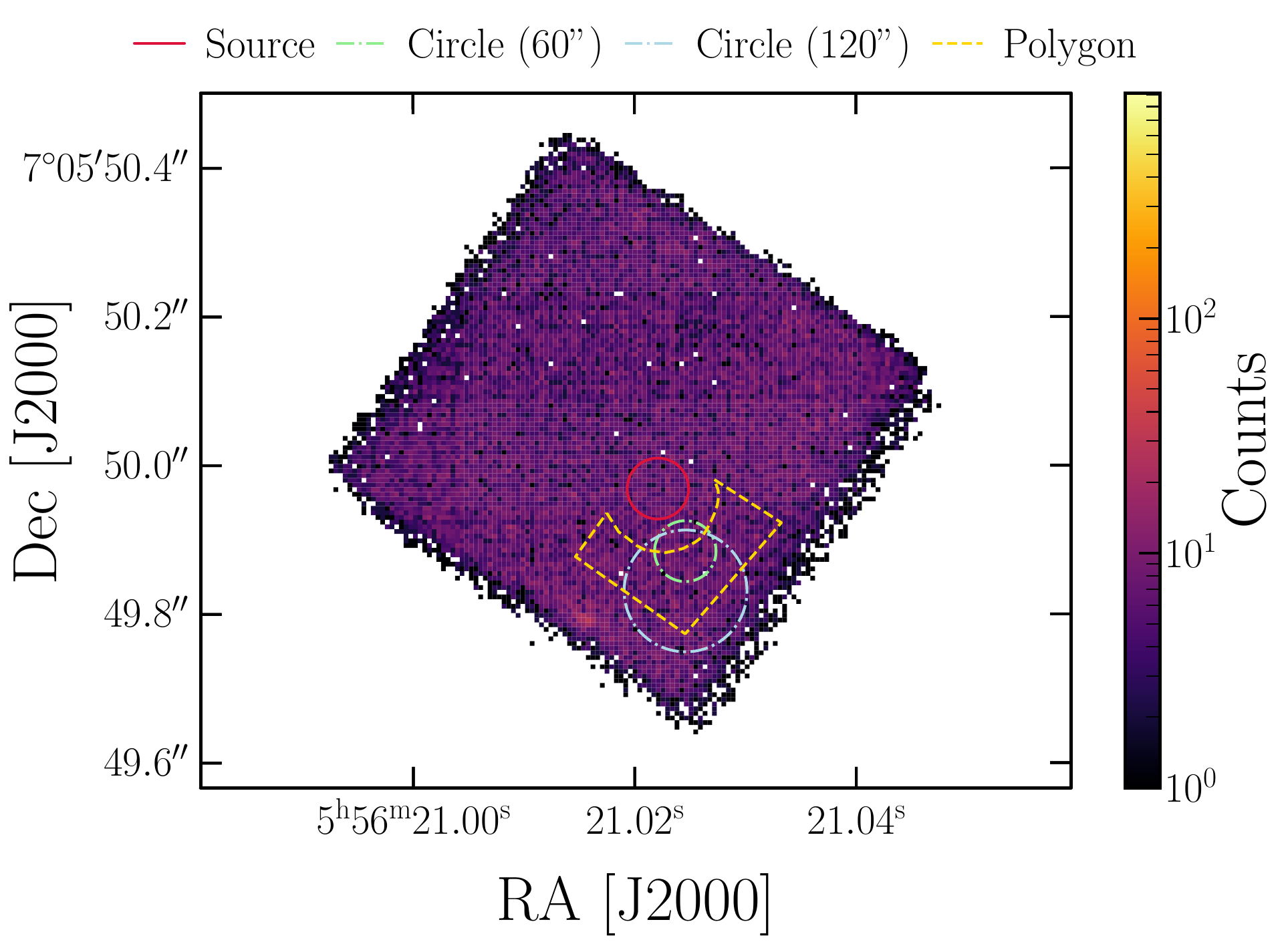}
    \label{fig:FOV_B}
    \end{subfigure}
      \caption{Field of view (FOV) of the Betelgeuse observation for module A (\textit{left}) and B (\textit{right}). The 60'' source region is displayed in a red solid line. On the other hand, the background regions considered in Appendix \ref{appx:background} are displayed as dashed lines. The polygon region, yellow dashed, is the background region used in the final analysis.}
    \label{fig:nustar_fov}
\end{figure*}

\textsc{NuSTAR}~\cite{NuSTAR:2013yza} is the first space-based focusing telescope capable of extending energy sensitivity into the hard x-ray regime, observing within the $3$--$79$~keV range. Launched into a near-equatorial low-Earth orbit in 2012, \textsc{NuSTAR} is equipped with two detection assemblies; focal plane modules A and B (FPMA/FPMB). Each module includes a telescope structure based on a segmented-glass design, featuring 133 nested shells coated with reflective multilayers. These layers consist of $\iso{}{Pt}/\iso{}{C}$ for the inner shells and $\iso{}{W}/\iso{}{Si}$ for the outer shells. The telescope employs a conical approximation of a Wolter-I geometry, optimized for grazing-incidence reflection. The focal plane detectors for each module consist of a $2 \times 2$ array of CdZnTe semiconductor chips. Each chip contains $32 \times 32$ pixels, resulting in a total of 4096 pixels per detector. Each pixel measures $0.6~\mathrm{mm} \times 0.6~\mathrm{mm}$, providing a FOV of approximately $12'$ in both $x$ and $y$ directions at $3$~keV. At higher energies ($>10$~keV), the FOV is typically reduced to between $6'$ and $10'$. The observatory achieves an angular resolution, determined mainly by its optics, of $18''$ full width at half maximum (FWHM) and a half-power diameter of about $60''$. Designed specifically for high performance in the hard x-ray spectrum, \textsc{NuSTAR} delivers an energy resolution with a FWHM of $0.4$~keV at $10$~keV and $0.9$~keV at $68$~keV.

The Betelgeuse observation examined in this research was retrieved from NASA's \texttt{HEASARC} data archive. To process the NuSTAR observation, we employed the NuSTARDAS software suite (version 2.1.4), which is part of HEASoft package (version 6.34) \cite{ascl:1408.004}. The workflow begins with reprocessing the raw data from the archival files using the \texttt{nupipeline} task provided by NuSTARDAS, with the flags \texttt{SAAMODE=OPTIMIZED} and \texttt{TENTACLE=YES} to remove intervals exhibiting elevated instrument backgrounds, typically occurring when the telescope passes through the south Atlantic anomaly. As explained in the main text and shown in Figure \ref{fig:nustar_fov}, the source region is defined as a 60'' circular region centered at Betelgeuse's coordinates. The background region is chosen to be a polygon separated 120'' from the source center and in the same CdZnTe detector chip as the source. For an extended discussion on the background region optimization, confer Appendix \ref{appx:background}. 

Next, \texttt{nuproducts}, with the previously mentioned regions, is used to subtract the source and normalized background spectra. 
Then, we employ HEASoft's \texttt{XSPEC} tool \cite{arnaud1999xspec} (version 12.14.1) to read the spectra and export the data into ASCII format.

On the other hand, the predicted differential photon flux from axion decay is forward modeled by incorporating the response files corresponding to both optics and detector, to determine the expected counts in the \textsc{NuSTAR} detector. The effective area of the telescope is taken into account by the ancillary response files (ARFs) and the detector energy response by the response matrix files (RMFs). Thus, our axion-induced expected counts for each energy bin $j$ and each module $i= \{\mathrm{A}, \mathrm{B}\}$ can be computed via 
\begin{widetext}
\begin{equation}
N_{\mathrm{ax},i}^{j}\left(m_a,g_{a\gamma}, g_{aN} \right)  =  \Delta t_i  
\int \mathrm{RMF}^i(E_j,E) \mathrm{ARF}^{i}\left(E\right)
\frac{\dd N_a(m_a, g_{aN})}{\dd E}P_{a \rightarrow \gamma}(E, g_{a\gamma}) \dd E \; , 
\end{equation}
\end{widetext}
where $\Delta t_i$ corresponds to the exposure time of the module $i$. 

\subsection{Background Region Optimization}\label{appx:background}

In order to accurately account for the spatial variation of the background components—primarily the unfocused cosmic x‐ray background and the x‐rays arising from fluorescence/activation of the instrument structure—across the detector arrays, we selected various background regions following the methodology of \cite{Xiao:2020pra}.
Given that the detector background intensity is known to vary between chips, while remaining largely constant within each chip \cite{Wik_2014}, we compared the background spectra from three distinct regions on a given chip: a 60'' region near the source, a 120'' region, and a polygonal region, each located at least 120'' away from the source. 

\begin{figure*}[t]
    \centering
    \begin{subfigure}[b]{0.49\textwidth}
        \centering
        \includegraphics[width=\textwidth]{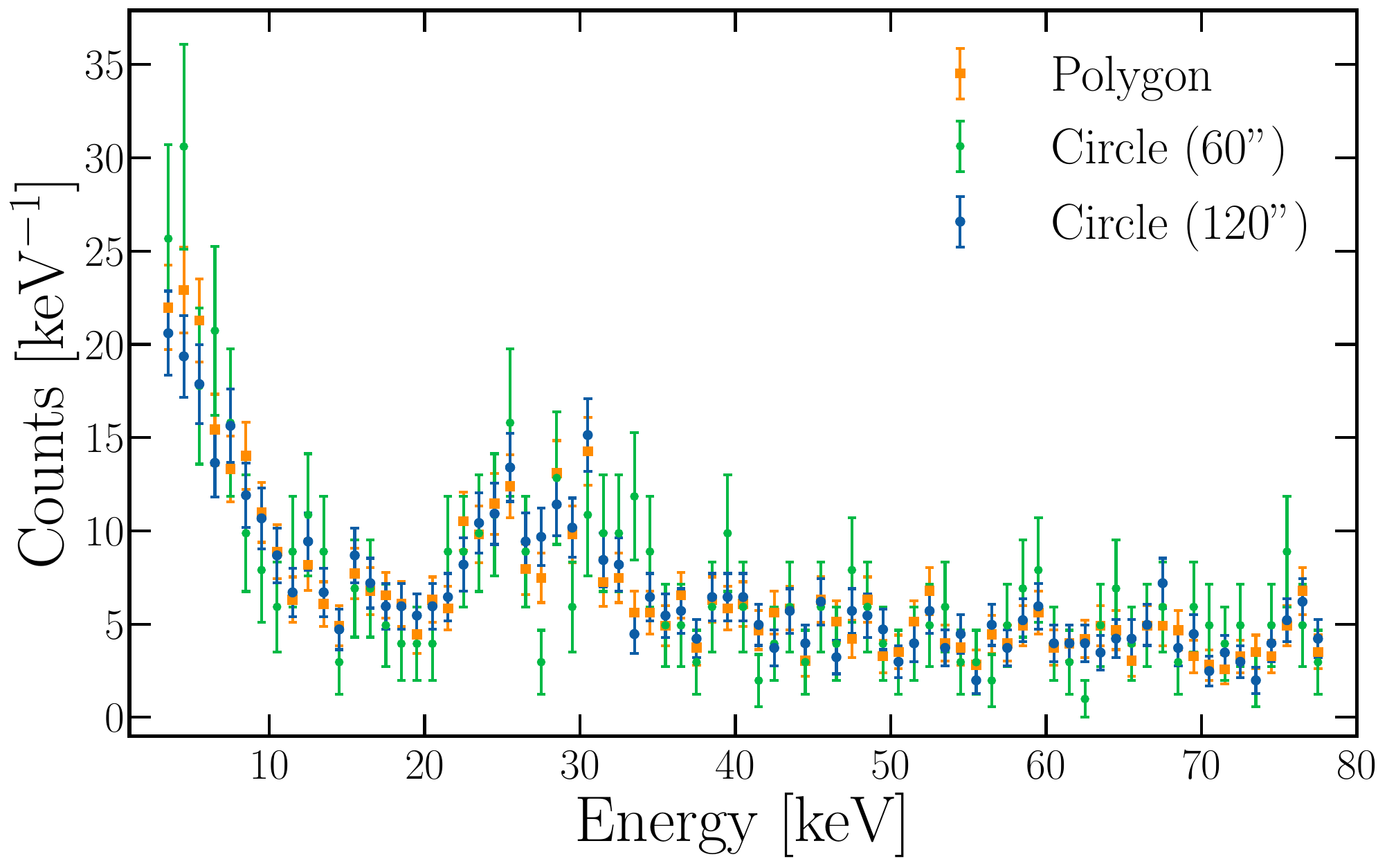}
        \label{fig:bg_A}
    \end{subfigure}
    \hfill
     \begin{subfigure}[b]{0.49\textwidth}
        \centering
        \includegraphics[width=\textwidth]{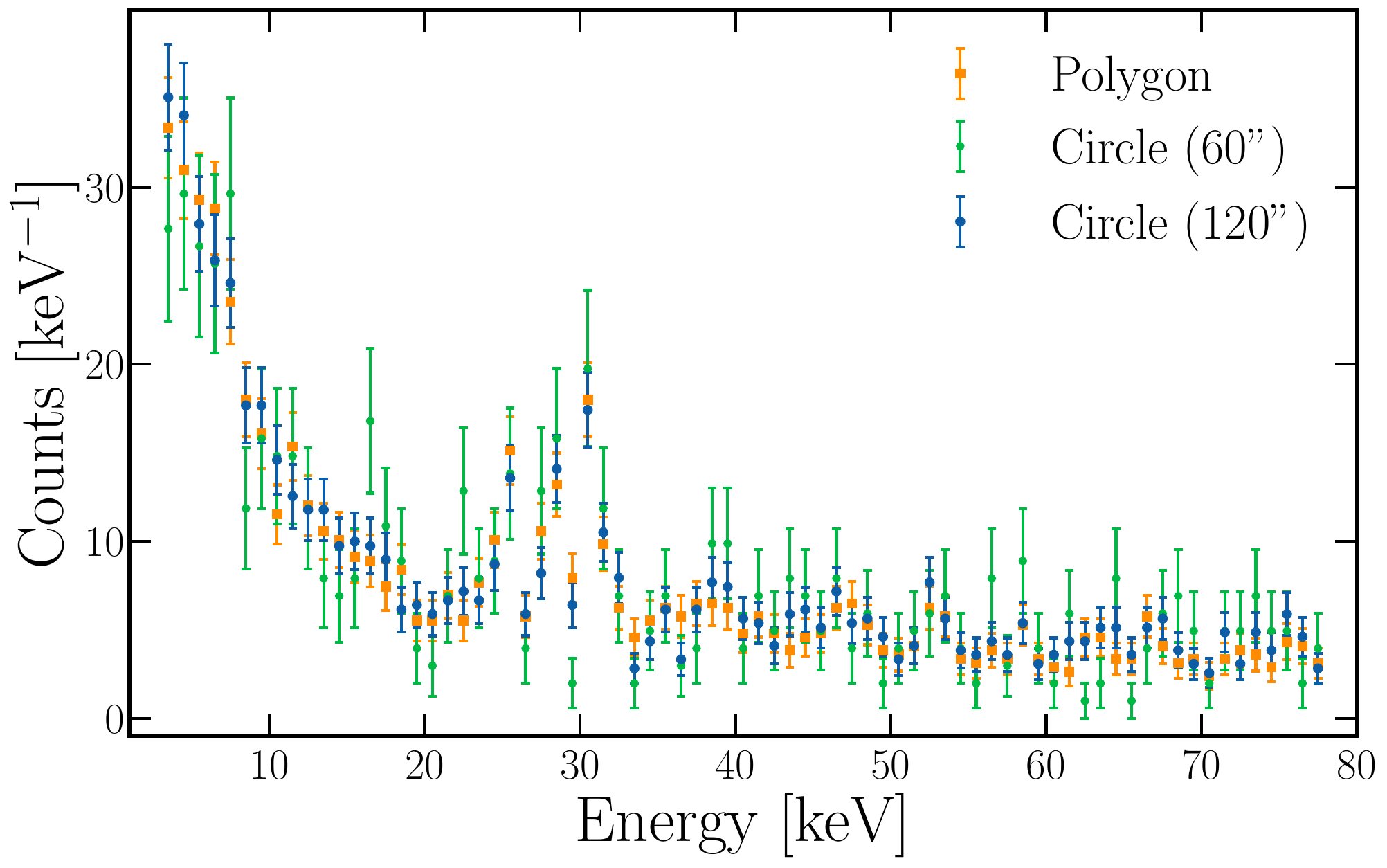}
        \label{fig:bg_B}
    \end{subfigure}
    \caption{X-ray spectra in the 3-79 keV energy range from FPMA (\textit{left}) and FPMB (\textit{right}) in different background regions after the normalization, the polygonal region in orange; the spectrum from the 60'' circle region in green and the spectrum from the 120'' circle region in blue. The data is binned in 1 keV increments and multiplied by the exposure time (49.160 ks for FPMA and 48.830 ks for FPMB).}
  \label{fig:nu_back_op}
\end{figure*}

To reinforce the selection of the polygonal background region, we calculated and compared the signal-to-noise ratio (SNR), defined as 
$
\operatorname{SNR} = \sum \textrm{Counts}/(\sum \textrm{Errors}^2)^{1/2},
$
for each region and compared the corresponding background spectra on the same detector chip; see Figure \ref{fig:nustar_fov} for a visual representation. The SNR values for each region are reported in Table \ref{tab:background_snr}, and the background spectra are shown in Figure \ref{fig:nu_back_op}.

\begin{table}[h]
\centering
\renewcommand{\arraystretch}{1.5} 
\setlength{\tabcolsep}{10pt} 
\caption{SNR values for the three background estimation regions in both Module A and Module B. The polygon region exhibits the highest SNR, making it the most reliable choice for background estimation.}
\begin{tabular}{lcc}
\toprule
Region & SNR (A) & SNR (B) \\
\hline
Polygon       & 47.00 & 50.01 \\
60'' Circle   & 23.30 & 25.08 \\
120'' Circle  & 45.50 & 49.10 \\
\hline
\hline
\end{tabular}

\label{tab:background_snr}
\end{table}
Although all regions exhibit similar background characteristics, the polygonal region yields the highest SNR. This indicates that the polygonal region provides a cleaner and more reliable background estimate with reduced noise contamination, thereby ensuring more precise background subtraction across both modules.  

\clearpage
\bibliography{fe57}

@article{Lella:2023bfb,
    author = "Lella, Alessandro and Carenza, Pierluca and Co', Giampaolo and Lucente, Giuseppe and Giannotti, Maurizio and Mirizzi, Alessandro and Rauscher, Thomas",
    title = "{Getting the most on supernova axions}",
    eprint = "2306.01048",
    archivePrefix = "arXiv",
    primaryClass = "hep-ph",
    doi = "10.1103/PhysRevD.109.023001",
    journal = "Phys. Rev. D",
    volume = "109",
    number = "2",
    pages = "023001",
    year = "2024"
}

@article{Lella:2022uwi,
    author = "Lella, Alessandro and Carenza, Pierluca and Lucente, Giuseppe and Giannotti, Maurizio and Mirizzi, Alessandro",
    title = "{Protoneutron stars as cosmic factories for massive axionlike particles}",
    eprint = "2211.13760",
    archivePrefix = "arXiv",
    primaryClass = "hep-ph",
    doi = "10.1103/PhysRevD.107.103017",
    journal = "Phys. Rev. D",
    volume = "107",
    number = "10",
    pages = "103017",
    year = "2023"
}

@article{Lella:2024hfk,
    author = "Lella, Alessandro and Calore, Francesca and Carenza, Pierluca and Eckner, Christopher and Giannotti, Maurizio and Lucente, Giuseppe and Mirizzi, Alessandro",
    title = "{Probing protoneutron stars with gamma-ray axionscopes}",
    eprint = "2405.02395",
    archivePrefix = "arXiv",
    primaryClass = "hep-ph",
    reportNumber = "LAPTH-024/24, BARI-TH/769-24",
    doi = "10.1088/1475-7516/2024/11/009",
    journal = "JCAP",
    volume = "11",
    pages = "009",
    year = "2024"
}

@article{Avignone:2017ylv,
	archiveprefix = {arXiv},
	author = {Avignone, F. T. and Creswick, R. J. and Vergados, J. D. and Pirinen, P. and Srivastava, P. C. and Suhonen, J.},
	doi = {10.1088/1475-7516/2018/01/021},
	eprint = {1711.06979},
	journal = {JCAP},
	pages = {021},
	primaryclass = {hep-ph},
	title = {{Estimating the flux of the 14.4 keV solar axions}},
	volume = {01},
	year = {2018}}

@misc{ascl:1408.004,
	adsurl = {https://ui.adsabs.harvard.edu/abs/2014ApJ...790...52M},
	author = {{NASA High Energy Astrophysics Science Archive Research Center (HEASARC)}},
	bibcode = {2014ascl.soft08004N},
	note = {{Astrophysics Source Code Library, record ascl:1408.004}},
	title = {{HEAsoft: Unified Release of FTOOLS and XANADU}},
	url = {https://heasarc.gsfc.nasa.gov/lheasoft/},
	year = 2014}

@article{Weinberg:1977ma,
	author = {Weinberg, Steven},
	doi = {10.1103/PhysRevLett.40.223},
	journal = {Phys. Rev. Lett.},
	pages = {223--226},
	reportnumber = {HUTP-77/A074},
	title = {{A New Light Boson?}},
	volume = {40},
	year = {1978}}

@article{Ning:2025tit,
    author = "Ning, Orion and Safdi, Benjamin R.",
    title = "{Probing the Axion-Electron Coupling with NuSTAR Observations of Galaxies}",
    eprint = "2503.09682",
    archivePrefix = "arXiv",
    primaryClass = "hep-ph",
    month = "3",
    year = "2025",
    journal = {arXiv},
}

@article{Ning:2024eky,
  title = {Leading Axion-Photon Sensitivity with NuSTAR Observations of M82 and M87},
  author = {Ning, Orion and Safdi, Benjamin R.},
  journal = {Phys. Rev. Lett.},
  volume = {134},
  issue = {17},
  pages = {171003},
  numpages = {7},
  year = {2025},
  month = {May},
  publisher = {American Physical Society},
  doi = {10.1103/PhysRevLett.134.171003},
  url = {https://link.aps.org/doi/10.1103/PhysRevLett.134.171003}
}

@article{Wilczek:1977pj,
	author = {Wilczek, Frank},
	doi = {10.1103/PhysRevLett.40.279},
	journal = {Phys. Rev. Lett.},
	pages = {279--282},
	reportnumber = {Print-77-0939 (COLUMBIA)},
	title = {{Problem of Strong $P$ and $T$ Invariance in the Presence of Instantons}},
	volume = {40},
	year = {1978}}

@article{Peccei:1977hh,
	author = {Peccei, R. D. and Quinn, Helen R.},
	doi = {10.1103/PhysRevLett.38.1440},
	journal = {Phys. Rev. Lett.},
	pages = {1440--1443},
	reportnumber = {ITP-568-STANFORD},
	title = {{CP Conservation in the Presence of Instantons}},
	volume = {38},
	year = {1977}}

@article{Peccei:1977ur,
	author = {Peccei, R. D. and Quinn, Helen R.},
	doi = {10.1103/PhysRevD.16.1791},
	journal = {Phys. Rev. D},
	pages = {1791--1797},
	reportnumber = {ITP-572-STANFORD},
	title = {{Constraints Imposed by CP Conservation in the Presence of Instantons}},
	volume = {16},
	year = {1977}}

@article{Witten:1984dg,
	author = {Witten, Edward},
	doi = {10.1016/0370-2693(84)90422-2},
	journal = {Phys. Lett. B},
	pages = {351--356},
	reportnumber = {Print-84-0838 (PRINCETON)},
	title = {{Some Properties of O(32) Superstrings}},
	volume = {149},
	year = {1984}}

@article{Conlon:2006tq,
	archiveprefix = {arXiv},
	author = {Conlon, Joseph P.},
	doi = {10.1088/1126-6708/2006/05/078},
	eprint = {hep-th/0602233},
	journal = {JHEP},
	pages = {078},
	reportnumber = {DAMTP-2006-17},
	title = {{The QCD axion and moduli stabilisation}},
	volume = {05},
	year = {2006}}

@article{Arvanitaki:2009fg,
	archiveprefix = {arXiv},
	author = {Arvanitaki, Asimina and Dimopoulos, Savas and Dubovsky, Sergei and Kaloper, Nemanja and March-Russell, John},
	doi = {10.1103/PhysRevD.81.123530},
	eprint = {0905.4720},
	journal = {Phys. Rev. D},
	pages = {123530},
	primaryclass = {hep-th},
	title = {{String Axiverse}},
	volume = {81},
	year = {2010}}

@article{Acharya:2010zx,
	archiveprefix = {arXiv},
	author = {Acharya, Bobby Samir and Bobkov, Konstantin and Kumar, Piyush},
	doi = {10.1007/JHEP11(2010)105},
	eprint = {1004.5138},
	journal = {JHEP},
	pages = {105},
	primaryclass = {hep-th},
	title = {{An M Theory Solution to the Strong CP Problem and Constraints on the Axiverse}},
	volume = {11},
	year = {2010}}

@article{Higaki:2011me,
	archiveprefix = {arXiv},
	author = {Higaki, Tetsutaro and Kobayashi, Tatsuo},
	doi = {10.1103/PhysRevD.84.045021},
	eprint = {1106.1293},
	journal = {Phys. Rev. D},
	pages = {045021},
	primaryclass = {hep-th},
	reportnumber = {DESY-11-094, KUNS-2340},
	title = {{Note on moduli stabilization, supersymmetry breaking and axiverse}},
	volume = {84},
	year = {2011}}

@article{Cicoli:2012sz,
	archiveprefix = {arXiv},
	author = {Cicoli, Michele and Goodsell, Mark and Ringwald, Andreas},
	doi = {10.1007/JHEP10(2012)146},
	eprint = {1206.0819},
	journal = {JHEP},
	pages = {146},
	primaryclass = {hep-th},
	reportnumber = {DESY-12-058, CERN-PH-TH-2012-153},
	title = {{The type IIB string axiverse and its low-energy phenomenology}},
	volume = {10},
	year = {2012}}

@article{Demirtas:2018akl,
	archiveprefix = {arXiv},
	author = {Demirtas, Mehmet and Long, Cody and McAllister, Liam and Stillman, Mike},
	doi = {10.1007/JHEP04(2020)138},
	eprint = {1808.01282},
	journal = {JHEP},
	pages = {138},
	primaryclass = {hep-th},
	title = {{The Kreuzer-Skarke Axiverse}},
	volume = {04},
	year = {2020}}

@article{Mehta:2021pwf,
	archiveprefix = {arXiv},
	author = {Mehta, Viraf M. and Demirtas, Mehmet and Long, Cody and Marsh, David J. E. and McAllister, Liam and Stott, Matthew J.},
	doi = {10.1088/1475-7516/2021/07/033},
	eprint = {2103.06812},
	journal = {JCAP},
	pages = {033},
	primaryclass = {hep-th},
	title = {{Superradiance in string theory}},
	volume = {07},
	year = {2021}}

@article{Irastorza:2018dyq,
	archiveprefix = {arXiv},
	author = {Irastorza, Igor G. and Redondo, Javier},
	doi = {10.1016/j.ppnp.2018.05.003},
	eprint = {1801.08127},
	journal = {Prog. Part. Nucl. Phys.},
	pages = {89--159},
	primaryclass = {hep-ph},
	title = {{New experimental approaches in the search for axion-like particles}},
	volume = {102},
	year = {2018}}

@article{DiVecchia:2019ejf,
	archiveprefix = {arXiv},
	author = {Di Vecchia, Paolo and Giannotti, Maurizio and Lattanzi, Massimiliano and Lindner, Axel},
	doi = {10.22323/1.336.0034},
	eprint = {1902.06567},
	journal = {PoS},
	pages = {034},
	primaryclass = {hep-ph},
	title = {{Round Table on Axions and Axion-like Particles}},
	volume = {Confinement2018},
	year = {2019}}

@article{DiLuzio:2020wdo,
	archiveprefix = {arXiv},
	author = {Di Luzio, Luca and Giannotti, Maurizio and Nardi, Enrico and Visinelli, Luca},
	doi = {10.1016/j.physrep.2020.06.002},
	eprint = {2003.01100},
	journal = {Phys. Rept.},
	pages = {1--117},
	primaryclass = {hep-ph},
	reportnumber = {DESY 20-036, DESY-20-036},
	title = {{The landscape of QCD axion models}},
	volume = {870},
	year = {2020}}

@article{Agrawal:2021dbo,
	archiveprefix = {arXiv},
	author = {Agrawal, Prateek and others},
	doi = {10.1140/epjc/s10052-021-09703-7},
	eprint = {2102.12143},
	journal = {Eur. Phys. J. C},
	number = {11},
	pages = {1015},
	primaryclass = {hep-ph},
	title = {{Feebly-interacting particles: FIPs 2020 workshop report}},
	volume = {81},
	year = {2021}}

@article{Sikivie:2020zpn,
	archiveprefix = {arXiv},
	author = {Sikivie, Pierre},
	doi = {10.1103/RevModPhys.93.015004},
	eprint = {2003.02206},
	journal = {Rev. Mod. Phys.},
	number = {1},
	pages = {015004},
	primaryclass = {hep-ph},
	title = {{Invisible Axion Search Methods}},
	volume = {93},
	year = {2021}}

@article{GrillidiCortona:2015jxo,
	archiveprefix = {arXiv},
	author = {Grilli di Cortona, Giovanni and Hardy, Edward and Pardo Vega, Javier and Villadoro, Giovanni},
	doi = {10.1007/JHEP01(2016)034},
	eprint = {1511.02867},
	journal = {JHEP},
	pages = {034},
	primaryclass = {hep-ph},
	title = {{The QCD axion, precisely}},
	volume = {01},
	year = {2016}}

@article{Keller:2012yr,
	archiveprefix = {arXiv},
	author = {Keller, Jochen and Sedrakian, Armen},
	doi = {10.1016/j.nuclphysa.2012.11.004},
	eprint = {1205.6940},
	journal = {Nucl. Phys. A},
	pages = {62--69},
	primaryclass = {astro-ph.CO},
	title = {{Axions from cooling compact stars}},
	volume = {897},
	year = {2013}}

@article{Sedrakian:2015krq,
	archiveprefix = {arXiv},
	author = {Sedrakian, Armen},
	doi = {10.1103/PhysRevD.93.065044},
	eprint = {1512.07828},
	journal = {Phys. Rev. D},
	number = {6},
	pages = {065044},
	primaryclass = {astro-ph.HE},
	title = {{Axion cooling of neutron stars}},
	volume = {93},
	year = {2016}}

@article{Hamaguchi:2018oqw,
	archiveprefix = {arXiv},
	author = {Hamaguchi, Koichi and Nagata, Natsumi and Yanagi, Keisuke and Zheng, Jiaming},
	doi = {10.1103/PhysRevD.98.103015},
	eprint = {1806.07151},
	journal = {Phys. Rev. D},
	number = {10},
	pages = {103015},
	primaryclass = {hep-ph},
	reportnumber = {UT-18-13, IPMU 18-0111, IPMU-18-0111},
	title = {{Limit on the Axion Decay Constant from the Cooling Neutron Star in Cassiopeia A}},
	volume = {98},
	year = {2018}}

@article{Beznogov:2018fda,
	archiveprefix = {arXiv},
	author = {Beznogov, Mikhail V. and Rrapaj, Ermal and Page, Dany and Reddy, Sanjay},
	doi = {10.1103/PhysRevC.98.035802},
	eprint = {1806.07991},
	journal = {Phys. Rev. C},
	number = {3},
	pages = {035802},
	primaryclass = {astro-ph.HE},
	reportnumber = {INT-PUB-18-029},
	title = {{Constraints on Axion-like Particles and Nucleon Pairing in Dense Matter from the Hot Neutron Star in HESS J1731-347}},
	volume = {98},
	year = {2018}}

@article{Sedrakian:2018kdm,
	archiveprefix = {arXiv},
	author = {Sedrakian, Armen},
	doi = {10.1103/PhysRevD.99.043011},
	eprint = {1810.00190},
	journal = {Phys. Rev. D},
	number = {4},
	pages = {043011},
	primaryclass = {astro-ph.HE},
	title = {{Axion cooling of neutron stars. II. Beyond hadronic axions}},
	volume = {99},
	year = {2019}}

@article{Leinson:2021ety,
	archiveprefix = {arXiv},
	author = {Leinson, Lev B.},
	doi = {10.1088/1475-7516/2021/09/001},
	eprint = {2105.14745},
	journal = {JCAP},
	pages = {001},
	primaryclass = {hep-ph},
	title = {{Impact of axions on the Cassiopea A neutron star cooling}},
	volume = {09},
	year = {2021}}

@article{Turner:1987by,
	author = {Turner, Michael S.},
	doi = {10.1103/PhysRevLett.60.1797},
	journal = {Phys. Rev. Lett.},
	pages = {1797},
	reportnumber = {FERMILAB-PUB-87-202-A},
	title = {{Axions from SN 1987a}},
	volume = {60},
	year = {1988}}

@article{Burrows:1988ah,
	author = {Burrows, Adam and Turner, Michael S. and Brinkmann, R. P.},
	doi = {10.1103/PhysRevD.39.1020},
	journal = {Phys. Rev. D},
	pages = {1020},
	reportnumber = {FERMILAB-PUB-88-105-A},
	title = {{Axions and SN 1987a}},
	volume = {39},
	year = {1989}}

@article{Raffelt:1987yt,
	author = {Raffelt, Georg and Seckel, David},
	doi = {10.1103/PhysRevLett.60.1793},
	journal = {Phys. Rev. Lett.},
	pages = {1793},
	reportnumber = {SCIPP-87/107},
	title = {{Bounds on Exotic Particle Interactions from SN 1987a}},
	volume = {60},
	year = {1988}}

@article{Raffelt:1990yz,
	author = {Raffelt, Georg G.},
	doi = {10.1016/0370-1573(90)90054-6},
	journal = {Phys. Rept.},
	pages = {1--113},
	reportnumber = {MPI-PAE-PTH-29-90},
	title = {{Astrophysical methods to constrain axions and other novel particle phenomena}},
	volume = {198},
	year = {1990}}

@article{Carenza:2019pxu,
	archiveprefix = {arXiv},
	author = {Carenza, Pierluca and Fischer, Tobias and Giannotti, Maurizio and Guo, Gang and Mart\'\i{}nez-Pinedo, Gabriel and Mirizzi, Alessandro},
	doi = {10.1088/1475-7516/2019/10/016},
	eprint = {1906.11844},
	journal = {JCAP},
	note = {[Erratum: JCAP 05, E01 (2020)]},
	number = {10},
	pages = {016},
	primaryclass = {hep-ph},
	title = {{Improved axion emissivity from a supernova via nucleon-nucleon bremsstrahlung}},
	volume = {10},
	year = {2019}}

@article{Carenza:2020cis,
	archiveprefix = {arXiv},
	author = {Carenza, Pierluca and Fore, Bryce and Giannotti, Maurizio and Mirizzi, Alessandro and Reddy, Sanjay},
	doi = {10.1103/PhysRevLett.126.071102},
	eprint = {2010.02943},
	journal = {Phys. Rev. Lett.},
	number = {7},
	pages = {071102},
	primaryclass = {hep-ph},
	reportnumber = {INT-PUB-20-039},
	title = {{Enhanced Supernova Axion Emission and its Implications}},
	volume = {126},
	year = {2021}}

@article{Fischer:2021jfm,
	archiveprefix = {arXiv},
	author = {Fischer, Tobias and Carenza, Pierluca and Fore, Bryce and Giannotti, Maurizio and Mirizzi, Alessandro and Reddy, Sanjay},
	doi = {10.1103/PhysRevD.104.103012},
	eprint = {2108.13726},
	journal = {Phys. Rev. D},
	number = {10},
	pages = {103012},
	primaryclass = {hep-ph},
	title = {{Observable signatures of enhanced axion emission from protoneutron stars}},
	volume = {104},
	year = {2021}}

@article{Donnelly:1978ty,
	author = {Donnelly, T. W. and Freedman, S. J. and Lytel, R. S. and Peccei, R. D. and Schwartz, M.},
	doi = {10.1103/PhysRevD.18.1607},
	journal = {Phys. Rev. D},
	pages = {1607},
	reportnumber = {ITP-598-STANFORD},
	title = {{Do Axions Exist?}},
	volume = {18},
	year = {1978}}

@article{CAST:2009jdc,
	archiveprefix = {arXiv},
	author = {Andriamonje, S. and others},
	collaboration = {CAST},
	doi = {10.1088/1475-7516/2009/12/002},
	eprint = {0906.4488},
	journal = {JCAP},
	pages = {002},
	primaryclass = {hep-ex},
	reportnumber = {SLAC-PUB-14825},
	title = {{Search for 14.4-keV solar axions emitted in the M1-transition of Fe-57 nuclei with CAST}},
	volume = {12},
	year = {2009}}

@article{XENON:2020rca,
	archiveprefix = {arXiv},
	author = {Aprile, E. and others},
	collaboration = {XENON},
	doi = {10.1103/PhysRevD.102.072004},
	eprint = {2006.09721},
	journal = {Phys. Rev. D},
	number = {7},
	pages = {072004},
	primaryclass = {hep-ex},
	title = {{Excess electronic recoil events in XENON1T}},
	volume = {102},
	year = {2020}}

@article{Xiao:2020pra,
	archiveprefix = {arXiv},
	author = {Xiao, Mengjiao and Perez, Kerstin M. and Giannotti, Maurizio and Straniero, Oscar and Mirizzi, Alessandro and Grefenstette, Brian W. and Roach, Brandon M. and Nynka, Melania},
	doi = {10.1103/PhysRevLett.126.031101},
	eprint = {2009.09059},
	journal = {Phys. Rev. Lett.},
	number = {3},
	pages = {031101},
	primaryclass = {astro-ph.HE},
	title = {{Constraints on Axionlike Particles from a Hard X-Ray Observation of Betelgeuse}},
	volume = {126},
	year = {2021}}

@article{Xiao:2022rxk,
	archiveprefix = {arXiv},
	author = {Xiao, Mengjiao and Carenza, Pierluca and Giannotti, Maurizio and Mirizzi, Alessandro and Perez, Kerstin M. and Straniero, Oscar and Grefenstette, Brian W.},
	doi = {10.1103/PhysRevD.106.123019},
	eprint = {2204.03121},
	journal = {Phys. Rev. D},
	number = {12},
	pages = {123019},
	primaryclass = {astro-ph.HE},
	title = {{Betelgeuse constraints on coupling between axionlike particles and electrons}},
	volume = {106},
	year = {2022}}

@article{NuSTAR:2013yza,
	archiveprefix = {arXiv},
	author = {Harrison, Fiona A. and others},
	collaboration = {NuSTAR},
	doi = {10.1088/0004-637X/770/2/103},
	eprint = {1301.7307},
	journal = {Astrophys. J.},
	pages = {103},
	primaryclass = {astro-ph.IM},
	reportnumber = {SLAC-PUB-16148},
	title = {{The Nuclear Spectroscopic Telescope Array (NuSTAR) High-Energy X-Ray Mission}},
	volume = {770},
	year = {2013}}

@book{Raffelt:1996wa,
	author = {Raffelt, G. G.},
	isbn = {978-0-226-70272-8},
	month = {5},
	title = {{Stars as laboratories for fundamental physics}: {The astrophysics of neutrinos, axions, and other weakly interacting particles}},
	year = {1996}}

@article{Giannotti:2015kwo,
	archiveprefix = {arXiv},
	author = {Giannotti, Maurizio and Irastorza, Igor and Redondo, Javier and Ringwald, Andreas},
	doi = {10.1088/1475-7516/2016/05/057},
	eprint = {1512.08108},
	journal = {JCAP},
	pages = {057},
	primaryclass = {astro-ph.HE},
	reportnumber = {DESY-15-245},
	title = {{Cool WISPs for stellar cooling excesses}},
	volume = {05},
	year = {2016}}

@article{Giannotti:2017hny,
	archiveprefix = {arXiv},
	author = {Giannotti, Maurizio and Irastorza, Igor G. and Redondo, Javier and Ringwald, Andreas and Saikawa, Ken'ichi},
	doi = {10.1088/1475-7516/2017/10/010},
	eprint = {1708.02111},
	journal = {JCAP},
	pages = {010},
	primaryclass = {hep-ph},
	reportnumber = {DESY-17-116},
	title = {{Stellar Recipes for Axion Hunters}},
	volume = {10},
	year = {2017}}

@article{DiLuzio:2020jjp,
	archiveprefix = {arXiv},
	author = {Di Luzio, Luca and Fedele, Marco and Giannotti, Maurizio and Mescia, Federico and Nardi, Enrico},
	doi = {10.1103/PhysRevLett.125.131804},
	eprint = {2006.12487},
	journal = {Phys. Rev. Lett.},
	number = {13},
	pages = {131804},
	primaryclass = {hep-ph},
	reportnumber = {DESY 20-106, DESY-20-106},
	title = {{Solar axions cannot explain the XENON1T excess}},
	volume = {125},
	year = {2020}}

@article{DiLuzio:2021ysg,
	archiveprefix = {arXiv},
	author = {Di Luzio, Luca and Fedele, Marco and Giannotti, Maurizio and Mescia, Federico and Nardi, Enrico},
	doi = {10.1088/1475-7516/2022/02/035},
	eprint = {2109.10368},
	journal = {JCAP},
	number = {02},
	pages = {035},
	primaryclass = {hep-ph},
	reportnumber = {DESY-21-141, TTP21-030, P3H-21-062},
	title = {{Stellar evolution confronts axion models}},
	volume = {02},
	year = {2022}}

@article{Caputo:2024oqc,
	archiveprefix = {arXiv},
	author = {Caputo, Andrea and Raffelt, Georg},
	doi = {10.22323/1.454.0041},
	eprint = {2401.13728},
	journal = {PoS},
	pages = {041},
	primaryclass = {hep-ph},
	reportnumber = {MPP-2024-13, CERN-TH-2024-013},
	title = {{Astrophysical Axion Bounds: The 2024 Edition}},
	volume = {COSMICWISPers},
	year = {2024}}

@article{Carenza:2024ehj,
    author = "Carenza, Pierluca and Giannotti, Maurizio and Isern, Jordi and Mirizzi, Alessandro and Straniero, Oscar",
    title = "{Axion astrophysics}",
    eprint = "2411.02492",
    archivePrefix = "arXiv",
    primaryClass = "hep-ph",
    reportNumber = "BARI-TH/66-24",
    doi = "10.1016/j.physrep.2025.02.002",
    journal = "Phys. Rept.",
    volume = "1117",
    pages = "1--102",
    year = "2025"
}

@article{Borexino:2012guz,
	archiveprefix = {arXiv},
	author = {Bellini, G. and others},
	collaboration = {Borexino},
	doi = {10.1103/PhysRevD.85.092003},
	eprint = {1203.6258},
	journal = {Phys. Rev. D},
	pages = {092003},
	primaryclass = {hep-ex},
	title = {{Search for Solar Axions Produced in $p(d,\rm{^3He})A$ Reaction with Borexino Detector}},
	volume = {85},
	year = {2012}}

@article{Bhusal:2020bvx,
	archiveprefix = {arXiv},
	author = {Bhusal, Aagaman and Houston, Nick and Li, Tianjun},
	doi = {10.1103/PhysRevLett.126.091601},
	eprint = {2004.02733},
	journal = {Phys. Rev. Lett.},
	number = {9},
	pages = {091601},
	primaryclass = {hep-ph},
	title = {{Searching for Solar Axions Using Data from the Sudbury Neutrino Observatory}},
	volume = {126},
	year = {2021}}

@article{Arias-Aragon:2024gdz,
    author = "Arias-Arag{\'o}n, Fernando and Giannotti, Maurizio and di Cortona, Giovanni Grilli and Mescia, Federico",
    title = "{Axion-induced pair production: A new strategy for axion detection}",
    eprint = "2411.19327",
    archivePrefix = "arXiv",
    primaryClass = "hep-ph",
    doi = "10.1103/PhysRevD.111.043021",
    journal = "Phys. Rev. D",
    volume = "111",
    number = "4",
    pages = "043021",
    year = "2025"
}

@article{CAST:2009klq,
	archiveprefix = {arXiv},
	author = {Andriamonje, S. and others},
	collaboration = {CAST},
	doi = {10.1088/1475-7516/2010/03/032},
	eprint = {0904.2103},
	journal = {JCAP},
	pages = {032},
	primaryclass = {hep-ex},
	reportnumber = {FERMILAB-PUB-09-854},
	title = {{Search for solar axion emission from $^7Li$ and $D(p,\gamma)^3He$ nuclear decays with the CAST $\gamma$-ray calorimeter}},
	volume = {03},
	year = {2010}}

@article{DiLuzio:2021qct,
	archiveprefix = {arXiv},
	author = {Di Luzio, Luca and others},
	doi = {10.1140/epjc/s10052-022-10061-1},
	eprint = {2111.06407},
	journal = {Eur. Phys. J. C},
	number = {2},
	pages = {120},
	primaryclass = {hep-ph},
	reportnumber = {DESY-21-194},
	title = {{Probing the axion\textendash{}nucleon coupling with the next generation of~axion helioscopes}},
	volume = {82},
	year = {2022}}

@article{Kervella_18,
	author = {{Kervella, Pierre} and {Decin, Leen} and {Richards, Anita M. S.} and {Harper, Graham M.} and {McDonald, Iain} and {O'Gorman, Eamon} and {Montarg{\`e}s, Miguel} and {Homan, Ward} and {Ohnaka, Keiichi}},
	doi = {10.1051/0004-6361/201731761},
	journal = {A\&A},
	pages = {A67},
	title = {The close circumstellar environment of Betelgeuse - V. Rotation velocity and molecular envelope properties from ALMA},
	url = {https://doi.org/10.1051/0004-6361/201731761},
	volume = 609,
	year = 2018}

@ARTICLE{XuHan_2019,
       author = {{Xu}, J. and {Han}, J.~L.},
        title = "{Magnetic fields in the solar vicinity and in the Galactic halo}",
      journal = {Monthly Notes of the Royal Astronomical Society},
         year = 2019,
        month = jul,
       volume = {486},
       number = {3},
        pages = {4275-4289},
          doi = {10.1093/mnras/stz1060},
       adsurl = {https://ui.adsabs.harvard.edu/abs/2019MNRAS.486.4275X}
}

@article{Wik_2014,
doi = {10.1088/0004-637X/792/1/48},
url = {https://dx.doi.org/10.1088/0004-637X/792/1/48},
year = {2014},
month = {aug},
publisher = {The American Astronomical Society},
volume = {792},
number = {1},
pages = {48},
author = {Wik, Daniel R. and Hornstrup, A. and Molendi, S. and Madejski, G. and Harrison, F. A. and Zoglauer, A. and Grefenstette, B. W. and Gastaldello, F. and Madsen, K. K. and Westergaard, N. J. and Ferreira, D. D. M. and Kitaguchi, T. and Pedersen, K. and Boggs, S. E. and Christensen, F. E. and Craig, W. W. and Hailey, C. J. and Stern, D. and Zhang, W. W.},
title = {NuSTAR OBSERVATIONS OF THE BULLET CLUSTER: CONSTRAINTS ON INVERSE COMPTON EMISSION},
journal = {The Astrophysical Journal}
}

@article{Ruz:2024gkl,
    author = "Ruz, J. and others",
    title = "{NuSTAR as an Axion Helioscope}",
    eprint = "2407.03828",
    archivePrefix = "arXiv",
    primaryClass = "astro-ph.CO",
    doi = "10.1103/18sn-hxtb",
    journal = "Phys. Rev. Lett.",
    volume = "135",
    number = "14",
    pages = "141001",
    year = "2025"
}

@article{Li:2015tyq,
    author = "Li, Dawei and Creswick, Richard J. and Avignone, Frank T. and Wang, Yuanxu",
    title = "{Sensitivity of the CUORE detector to $14.4$ keV solar axions emitted by the M1 nuclear transition of$~^{57}$Fe}",
    eprint = "1512.01298",
    archivePrefix = "arXiv",
    primaryClass = "astro-ph.CO",
    doi = "10.1088/1475-7516/2016/02/031",
    journal = "JCAP",
    volume = "02",
    pages = "031",
    year = "2016"
}

@article{CAST:2024eil,
    author = {Altenm\"uller, K. and others},
    collaboration = "CAST",
    title = "{New Upper Limit on the Axion-Photon Coupling with an Extended CAST Run with a Xe-Based Micromegas Detector}",
    eprint = "2406.16840",
    archivePrefix = "arXiv",
    primaryClass = "hep-ex",
    doi = "10.1103/PhysRevLett.133.221005",
    journal = "Phys. Rev. Lett.",
    volume = "133",
    number = "22",
    pages = "221005",
    year = "2024"
}

@article{Carenza:2021alz,
         author = "Carenza, Pierluca and Evoli, Carmelo and Giannotti, Maurizio and Mirizzi, Alessandro and Montanino, Daniele",
    title = "{Turbulent axion-photon conversions in the Milky~Way}",
    eprint = "2104.13935",
    archivePrefix = "arXiv",
    primaryClass = "hep-ph",
    doi = "10.1103/PhysRevD.104.023003",
    journal = "Phys. Rev. D",
    volume = "104",
    number = "2",
    pages = "023003",
    year = "2021"
}

@article{OShea:2023gqn,
    author = "O'Shea, T. and Giannotti, M. and Irastorza, I. G. and Plasencia, L. M. and Redondo, J. and Ruz, J. and Vogel, J. K.",
    title = "{Prospects on the detection of solar dark photons by the International Axion Observatory}",
    eprint = "2312.10150",
    archivePrefix = "arXiv",
    primaryClass = "hep-ph",
    doi = "10.1088/1475-7516/2024/06/070",
    journal = "JCAP",
    volume = "06",
    pages = "070",
    year = "2024"
}

@article{Payez:2014xsa,
    author = "Payez, Alexandre and Evoli, Carmelo and Fischer, Tobias and Giannotti, Maurizio and Mirizzi, Alessandro and Ringwald, Andreas",
    title = "{Revisiting the SN1987A gamma-ray limit on ultralight axion-like particles}",
    eprint = "1410.3747",
    archivePrefix = "arXiv",
    primaryClass = "astro-ph.HE",
    reportNumber = "DESY-14-164",
    doi = "10.1088/1475-7516/2015/02/006",
    journal = "JCAP",
    volume = "02",
    pages = "006",
    year = "2015"
}

@article{Hoof:2022xbe,
    author = "Hoof, Sebastian and Schulz, Lena",
    title = "{Updated constraints on axion-like particles from temporal information in supernova SN1987A gamma-ray data}",
    eprint = "2212.09764",
    archivePrefix = "arXiv",
    primaryClass = "hep-ph",
    reportNumber = "TTP22-072",
    doi = "10.1088/1475-7516/2023/03/054",
    journal = "JCAP",
    volume = "03",
    pages = "054",
    year = "2023"
}

@article{Calore:2020tjw,
    author = "Calore, Francesca and Carenza, Pierluca and Giannotti, Maurizio and Jaeckel, Joerg and Mirizzi, Alessandro",
    title = "{Bounds on axionlike particles from the diffuse supernova flux}",
    eprint = "2008.11741",
    archivePrefix = "arXiv",
    primaryClass = "hep-ph",
    doi = "10.1103/PhysRevD.102.123005",
    journal = "Phys. Rev. D",
    volume = "102",
    number = "12",
    pages = "123005",
    year = "2020"
}

@article{Lucente:2022esm,
    author = "Lucente, Giuseppe and Nath, Newton and Capozzi, Francesco and Giannotti, Maurizio and Mirizzi, Alessandro",
    title = "{Probing high-energy solar axion flux with a large scintillation neutrino detector}",
    eprint = "2209.11780",
    archivePrefix = "arXiv",
    primaryClass = "hep-ph",
    doi = "10.1103/PhysRevD.106.123007",
    journal = "Phys. Rev. D",
    volume = "106",
    number = "12",
    pages = "123007",
    year = "2022"
}

@article{Beck:2012,
	author = {Beck, Paul G. and Montalban, Josefina and Kallinger, Thomas and De Ridder, Joris and Aerts, Conny and Garc{\'\i}a, Rafael A. and Hekker, Saskia and Dupret, Marc-Antoine and Mosser, Benoit and Eggenberger, Patrick and Stello, Dennis and Elsworth, Yvonne and Frandsen, S{\o}ren and Carrier, Fabien and Hillen, Michel and Gruberbauer, Michael and Christensen-Dalsgaard, J{\o}rgen and Miglio, Andrea and Valentini, Marica and Bedding, Timothy R. and Kjeldsen, Hans and Girouard, Forrest R. and Hall, Jennifer R. and Ibrahim, Khadeejah A.},
	date = {2012/01/01},
	doi = {10.1038/nature10612},
	id = {Beck2012},
	isbn = {1476-4687},
	journal = {Nature},
	number = {7379},
	pages = {55--57},
	title = {Fast core rotation in red-giant stars as revealed by gravity-dominated mixed modes},
	url = {https://doi.org/10.1038/nature10612},
	volume = {481},
	year = {2012}}

@ARTICLE{MESA_orig,
       author = {{Paxton}, Bill and {Bildsten}, Lars and {Dotter}, Aaron and {Herwig}, Falk and {Lesaffre}, Pierre and {Timmes}, Frank},
        title = "{Modules for Experiments in Stellar Astrophysics (MESA)}",
      journal = {Astrophysical Journal Supplement Series},
         year = 2011,
        month = jan,
       volume = {192},
       number = {1},
          eid = {3},
        pages = {3},
          doi = {10.1088/0067-0049/192/1/3},
archivePrefix = {arXiv},
       eprint = {1009.1622},
 primaryClass = {astro-ph.SR},
       adsurl = {https://ui.adsabs.harvard.edu/abs/2011ApJS..192....3P}
}

@ARTICLE{MESA2013,
       author = {{Paxton}, Bill and {Cantiello}, Matteo and {Arras}, Phil and {Bildsten}, Lars and {Brown}, Edward F. and {Dotter}, Aaron and {Mankovich}, Christopher and {Montgomery}, M.~H. and {Stello}, Dennis and {Timmes}, F.~X. and {Townsend}, Richard},
        title = "{Modules for Experiments in Stellar Astrophysics (MESA): Planets, Oscillations, Rotation, and Massive Stars}",
      journal = {The Astrophysical Journal Supplement Series},
         year = 2013,
        month = sep,
       volume = {208},
       number = {1},
          eid = {4},
        pages = {4},
          doi = {10.1088/0067-0049/208/1/4},
archivePrefix = {arXiv},
       eprint = {1301.0319},
 primaryClass = {astro-ph.SR},
       adsurl = {https://ui.adsabs.harvard.edu/abs/2013ApJS..208....4P}
}

@ARTICLE{MESA2015,
       author = {{Paxton}, Bill and {Marchant}, Pablo and {Schwab}, Josiah and {Bauer}, Evan B. and {Bildsten}, Lars and {Cantiello}, Matteo and {Dessart}, Luc and {Farmer}, R. and {Hu}, H. and {Langer}, N. and {Townsend}, R.~H.~D. and {Townsley}, Dean M. and {Timmes}, F.~X.},
        title = "{Modules for Experiments in Stellar Astrophysics (MESA): Binaries, Pulsations, and Explosions}",
      journal = {The Astrophysical Journal Supplement Series},
         year = 2015,
        month = sep,
       volume = {220},
       number = {1},
          eid = {15},
        pages = {15},
          doi = {10.1088/0067-0049/220/1/15},
archivePrefix = {arXiv},
       eprint = {1506.03146},
 primaryClass = {astro-ph.SR},
       adsurl = {https://ui.adsabs.harvard.edu/abs/2015ApJS..220...15P}
}

@ARTICLE{MESA2018,
       author = {{Paxton}, Bill and {Schwab}, Josiah and {Bauer}, Evan B. and {Bildsten}, Lars and {Blinnikov}, Sergei and {Duffell}, Paul and {Farmer}, R. and {Goldberg}, Jared A. and {Marchant}, Pablo and {Sorokina}, Elena and {Thoul}, Anne and {Townsend}, Richard H.~D. and {Timmes}, F.~X.},
        title = "{Modules for Experiments in Stellar Astrophysics (MESA): Convective Boundaries, Element Diffusion, and Massive Star Explosions}",
      journal = {The Astrophysical Journal Supplement Series},
         year = 2018,
        month = feb,
       volume = {234},
       number = {2},
          eid = {34},
        pages = {34},
          doi = {10.3847/1538-4365/aaa5a8},
archivePrefix = {arXiv},
       eprint = {1710.08424},
 primaryClass = {astro-ph.SR},
       adsurl = {https://ui.adsabs.harvard.edu/abs/2018ApJS..234...34P}
}

@ARTICLE{MESA2019,
       author = {{Paxton}, Bill and {Smolec}, R. and {Schwab}, Josiah and {Gautschy}, A. and {Bildsten}, Lars and {Cantiello}, Matteo and {Dotter}, Aaron and {Farmer}, R. and {Goldberg}, Jared A. and {Jermyn}, Adam S. and {Kanbur}, S.~M. and {Marchant}, Pablo and {Thoul}, Anne and {Townsend}, Richard H.~D. and {Wolf}, William M. and {Zhang}, Michael and {Timmes}, F.~X.},
        title = "{Modules for Experiments in Stellar Astrophysics (MESA): Pulsating Variable Stars, Rotation, Convective Boundaries, and Energy Conservation}",
      journal = {The Astrophysical Journal Supplement Series},
         year = 2019,
        month = jul,
       volume = {243},
       number = {1},
          eid = {10},
        pages = {10},
          doi = {10.3847/1538-4365/ab2241},
archivePrefix = {arXiv},
       eprint = {1903.01426},
 primaryClass = {astro-ph.SR},
       adsurl = {https://ui.adsabs.harvard.edu/abs/2019ApJS..243...10P}
}

@ARTICLE{MESA2023,
       author = {{Jermyn}, Adam S. and {Bauer}, Evan B. and {Schwab}, Josiah and {Farmer}, R. and {Ball}, Warrick H. and {Bellinger}, Earl P. and {Dotter}, Aaron and {Joyce}, Meridith and {Marchant}, Pablo and {Mombarg}, Joey S.~G. and {Wolf}, William M. and {Sunny Wong}, Tin Long and {Cinquegrana}, Giulia C. and {Farrell}, Eoin and {Smolec}, R. and {Thoul}, Anne and {Cantiello}, Matteo and {Herwig}, Falk and {Toloza}, Odette and {Bildsten}, Lars and {Townsend}, Richard H.~D. and {Timmes}, F.~X.},
        title = "{Modules for Experiments in Stellar Astrophysics (MESA): Time-dependent Convection, Energy Conservation, Automatic Differentiation, and Infrastructure}",
      journal = {The Astrophysical Journal Supplement Series},
         year = 2023,
        month = mar,
       volume = {265},
       number = {1},
          eid = {15},
        pages = {15},
          doi = {10.3847/1538-4365/acae8d},
archivePrefix = {arXiv},
       eprint = {2208.03651},
 primaryClass = {astro-ph.SR},
       adsurl = {https://ui.adsabs.harvard.edu/abs/2023ApJS..265...15J}
}

@article{Giannotti:2024xhx,
    author = "Giannotti, Maurizio",
    title = "{Status and Perspectives on Axion searches}",
    eprint = "2412.08733",
    archivePrefix = "arXiv",
    primaryClass = "hep-ph",
    doi = "10.22323/1.474.0033",
    journal = "PoS",
    volume = "COSMICWISPers2024",
    pages = "033",
    year = "2025"
}

@misc{cordes2003ne2001inewmodelgalactic,
      title={NE2001.I. A New Model for the Galactic Distribution of Free Electrons and its Fluctuations}, 
      author={J. M. Cordes and T. J. W. Lazio},
      year={2003},
      eprint={astro-ph/0207156},
      archivePrefix={arXiv},
      primaryClass={astro-ph},
      url={https://arxiv.org/abs/astro-ph/0207156}, 
}

@article{Harvey_Smith_2011,
   title={MAGNETIC FIELDS IN LARGE-DIAMETER H II REGIONS REVEALED BY THE FARADAY ROTATION OF COMPACT EXTRAGALACTIC RADIO SOURCES},
   volume={736},
   ISSN={1538-4357},
   url={http://dx.doi.org/10.1088/0004-637X/736/2/83},
   DOI={10.1088/0004-637x/736/2/83},
   number={2},
   journal={The Astrophysical Journal},
   publisher={American Astronomical Society},
   author={Harvey-Smith, L. and Madsen, G. J. and Gaensler, B. M.},
   year={2011},
   month=jul, pages={83} }

@article{arnaud1999xspec,
  title={XSPEC: An X-ray spectral fitting package},
  author={Arnaud, Keith and et al.},
  journal={Astrophysics Source Code Library},
  pages={ascl--9910},
  year={1999}
}

@article{Candon:2024eah,
  title = {NuSTAR Bounds on Radiatively Decaying Particles from M82},
  author = {Cand\'on, Francisco R. and Fiorillo, Damiano F. G. and Lucente, Giuseppe and Vitagliano, Edoardo and Vogel, Julia K.},
  journal = {Phys. Rev. Lett.},
  volume = {134},
  issue = {17},
  pages = {171004},
  numpages = {7},
  year = {2025},
  month = {May},
  publisher = {American Physical Society},
  doi = {10.1103/PhysRevLett.134.171004},
  url = {https://link.aps.org/doi/10.1103/PhysRevLett.134.171004}
}

@article{Meyer:2021pbp,
    author = "Meyer, Manuel and Davies, James and Kuhlmann, Julian",
    title = "{gammaALPs: An open-source python package for computing photon-axion-like-particle oscillations in astrophysical environments}",
    eprint = "2108.02061",
    archivePrefix = "arXiv",
    primaryClass = "astro-ph.HE",
    doi = "10.22323/1.395.0557",
    journal = "PoS",
    volume = "ICRC2021",
    pages = "557",
    year = "2021"
}

@article{Meyer:2014epa,
    author = "Meyer, Manuel and Montanino, Daniele and Conrad, Jan",
    title = "{On detecting oscillations of gamma rays into axion-like particles in turbulent and coherent magnetic fields}",
    eprint = "1406.5972",
    archivePrefix = "arXiv",
    primaryClass = "astro-ph.HE",
    doi = "10.1088/1475-7516/2014/09/003",
    journal = "JCAP",
    volume = "09",
    pages = "003",
    year = "2014"
}

@article{Darme:2020gyx,
    author = "Darm{\'e}, Luc and Di Luzio, Luca and Giannotti, Maurizio and Nardi, Enrico",
    title = "{Selective enhancement of the QCD axion couplings}",
    eprint = "2010.15846",
    archivePrefix = "arXiv",
    primaryClass = "hep-ph",
    reportNumber = "DESY-20-177, DESY 20-177",
    doi = "10.1103/PhysRevD.103.015034",
    journal = "Phys. Rev. D",
    volume = "103",
    number = "1",
    pages = "015034",
    year = "2021"
}

@article{Dessert:2020lil,
    author = "Dessert, Christopher and Foster, Joshua W. and Safdi, Benjamin R.",
    title = "{X-ray Searches for Axions from Super Star Clusters}",
    eprint = "2008.03305",
    archivePrefix = "arXiv",
    primaryClass = "hep-ph",
    doi = "10.1103/PhysRevLett.125.261102",
    journal = "Phys. Rev. Lett.",
    volume = "125",
    number = "26",
    pages = "261102",
    year = "2020"
}

@article{Buschmann:2021juv,
    author = "Buschmann, Malte and Dessert, Christopher and Foster, Joshua W. and Long, Andrew J. and Safdi, Benjamin R.",
    title = "{Upper Limit on the QCD Axion Mass from Isolated Neutron Star Cooling}",
    eprint = "2111.09892",
    archivePrefix = "arXiv",
    primaryClass = "hep-ph",
    doi = "10.1103/PhysRevLett.128.091102",
    journal = "Phys. Rev. Lett.",
    volume = "128",
    number = "9",
    pages = "091102",
    year = "2022"
}

@article{IAXO:2024wss,
    author = "Ahyoune, S. and others",
    collaboration = "IAXO",
    title = "{An accurate solar axions ray-tracing response of BabyIAXO}",
    eprint = "2411.13915",
    archivePrefix = "arXiv",
    primaryClass = "hep-ex",
    doi = "10.1007/JHEP02(2025)159",
    journal = "JHEP",
    volume = "02",
    pages = "159",
    year = "2025"
}

@article{IAXO:2020wwp,
    author = "Abeln, A. and others",
    collaboration = "IAXO",
    title = "{Conceptual design of BabyIAXO, the intermediate stage towards the International Axion Observatory}",
    eprint = "2010.12076",
    archivePrefix = "arXiv",
    primaryClass = "physics.ins-det",
    doi = "10.1007/JHEP05(2021)137",
    journal = "JHEP",
    volume = "05",
    pages = "137",
    year = "2021"
}

@article{IAXO:2019mpb,
    author = "Armengaud, E. and others",
    collaboration = "IAXO",
    title = "{Physics potential of the International Axion Observatory (IAXO)}",
    eprint = "1904.09155",
    archivePrefix = "arXiv",
    primaryClass = "hep-ph",
    doi = "10.1088/1475-7516/2019/06/047",
    journal = "JCAP",
    volume = "06",
    pages = "047",
    year = "2019"
}

@article{Manzari:2024jns,
    author = "Manzari, Claudio Andrea and Park, Yujin and Safdi, Benjamin R. and Savoray, Inbar",
    title = "{Supernova Axions Convert to Gamma Rays in Magnetic Fields of Progenitor Stars}",
    eprint = "2405.19393",
    archivePrefix = "arXiv",
    primaryClass = "hep-ph",
    doi = "10.1103/PhysRevLett.133.211002",
    journal = "Phys. Rev. Lett.",
    volume = "133",
    number = "21",
    pages = "211002",
    year = "2024"
}

@article{Auriere:2010cb,
    author = "Auriere, M. and Donati, J. -F. and Konstantinova-Antova, R. and Perrin, G. and Petit, P. and Roudier, T.",
    title = "{The magnetic field of Betelgeuse: a local dynamo from giant convection cells?}",
    eprint = "1005.4845",
    archivePrefix = "arXiv",
    primaryClass = "astro-ph.SR",
    doi = "10.1051/0004-6361/201014925",
    journal = "Astron. Astrophys.",
    volume = "516",
    pages = "L2",
    year = "2010"
}

@misc{NuSTAR_ObsID30501012002,
  author       = "{NuSTAR Mission}",
  title        = "{NuSTAR observation data for Betelgeuse (ObsID 30501012002), Observation Date 2019-08-23, NASA High Energy Astrophysics Science Archive Research Center (HEASARC), NASA Goddard Space Flight Center}",
  year         = {2019},
  note  = "\url{https://heasarc.gsfc.nasa.gov/FTP/nustar/data/obs/05/3//30501012002/}"
}

\end{document}